
\input harvmac
\input amssym.def
\input epsf
\noblackbox

\newcount\figno
\figno=0
\def\fig#1#2#3{
\par\begingroup\parindent=0pt\leftskip=1cm\rightskip=1cm\parindent=0pt
\baselineskip=11pt
\global\advance\figno by 1
\midinsert
\epsfxsize=#3
\centerline{\epsfbox{#2}}
\vskip 12pt
\centerline{{\bf Figure \the\figno} #1}\par
\endinsert\endgroup\par}
\def\figlabel#1{\xdef#1{\the\figno}}

\def\pmb#1{\setbox0=\hbox{#1}%
 \kern-.025em\copy0\kern-\wd0
 \kern.05em\copy0\kern-\wd0
 \kern-.025em\raise.0433em\box0 }
\font\cmss=cmss10
\font\cmsss=cmss10 at 7pt
\def\rlx{\relax\leavevmode}
\def\Cop{\relax{\rm l\kern-.49em C}}
\def\CCop{\relax{\rm l\kern-.49em C}}
\def\Rop{\relax{\rm I\kern-.18em R}}
\def\Nop{\relax{\rm I\kern-.18em N}}
\def\Pop{\relax{\rm I\kern-.18em P}}
\def\Zop{\rlx\leavevmode\ifmmode\mathchoice{\hbox{\cmss Z\kern-.4em Z}}
 {\hbox{\cmss Z\kern-.4em Z}}{\lower.9pt\hbox{\cmsss Z\kern-.36em Z}}
 {\lower1.2pt\hbox{\cmsss Z\kern-.36em Z}}\else{\cmss Z\kern-.4em
 Z}\fi}


\def\Tr{\mbox{Tr}}
\def\STr{\hbox{STr}}
\def\ie{{\it i.e.}}
\def\eg{{\it e.g.}}

\def\IZ{\relax\ifmmode\mathchoice {\hbox{\cmss Z\kern-.4em
Z}}{\hbox{\cmss Z\kern-.4em Z}} {\lower.9pt\hbox{\cmsss Z\kern-.4em
Z}} {\lower1.2pt\hbox{\cmsss Z\kern-.4em Z}}\else{\cmss Z\kern-.4em
Z}\fi}

\def\Tr{{\rm Tr}}

\def\p{{\bf p}}

\def\half{{1\over 2}}

\parindent 25pt
\overfullrule=0pt
\tolerance=10000


\def\pmb#1{\setbox0=\hbox{#1}%
 \kern-.025em\copy0\kern-\wd0
 \kern.05em\copy0\kern-\wd0
 \kern-.025em\raise.0433em\box0 }
\font\cmss=cmss10
\font\cmsss=cmss10 at 7pt
\def\rlx{\relax\leavevmode}
\def\Zop{\rlx\leavevmode\ifmmode\mathchoice{\hbox{\cmss Z\kern-.4em Z}}
 {\hbox{\cmss Z\kern-.4em Z}}{\lower.9pt\hbox{\cmsss Z\kern-.36em Z}}
 {\lower1.2pt\hbox{\cmsss Z\kern-.36em Z}}\else{\cmss Z\kern-.4em
 Z}\fi}


\def\pmb#1{\setbox0=\hbox{#1}%
 \kern-.025em\copy0\kern-\wd0
 \kern.05em\copy0\kern-\wd0
 \kern-.025em\raise.0433em\box0 }
\font\cmss=cmss10
\font\cmsss=cmss10 at 7pt
\def\rlx{\relax\leavevmode}
\def\Cop{\relax{\rm C\kern-.58em C}}
\def\Rop{\relax{\rm I\kern-.18em R}}
\def\Nop{\relax{\rm I\kern-.18em N}}
\def\Pop{\relax{\rm I\kern-.18em P}}

\def\Zop{\rlx\leavevmode\ifmmode\mathchoice{\hbox{\cmss Z\kern-.4em Z}}
 {\hbox{\cmss Z\kern-.4em Z}}{\lower.9pt\hbox{\cmsss Z\kern-.36em Z}}
 {\lower1.2pt\hbox{\cmsss Z\kern-.36em Z}}\else{\cmss Z\kern-.4em
 Z}\fi}

\def\cf{{\it cf.}}
\def\slashint{\relax{\rm \hbox{-}\hbox{-} \kern-1.05em \int}}
\def\Slashint{\relax{\rm - \kern-.9em \int}}
\def\Gslash{ \, {\relax{/ \kern-.54em  G}}}
\def\Hslash{ \, {\relax{/ \kern-.66em  H}}}
\def\pslash{ \, {\relax{/ \kern-.55em  p}}}
\def\qslash{ \, {\relax{/ \kern-.55em  q}}}

\def\g{{\frak g}}
\def\psu{{\frak p}{\frak s}{\frak u}}
\def\su{{\frak s}{\frak u}}
\def\so{{\frak s}{\frak o}}
\def\gl{{\frak g}{\frak l}}
\def\ufrak{{\frak u}}
\def\slfrak{{\frak s}{\frak l}}
\def\C{{\cal C}}
\def\fourfour{{\bf 4}|{\bf 4}}
\def\Gt{\tilde{G}}
\def\rank{\hbox{rank}}
\def\R{{\cal R}}
\def\V{{\cal V}}
\def\T{{\cal T}}
\def\g{{\frak g}}


\lref\BeisertI{
N.~Beisert,
{\it The complete one-loop dilatation operator of N = 4 super
Yang-Mills theory},
Nucl.\ Phys.\ B {\bf 676}, 3 (2004);
{\tt hep-th/0307015}.}

\lref\BFHZ{
N.~Beisert, G.~Ferretti, R.~Heise and K.~Zarembo,
{\it One-Loop QCD Spin Chain and its Spectrum};
{\tt hep-th/0412029}.
}

\lref\Frolov{
S.~Frolov and A.~A.~Tseytlin,
{\it Multi-spin string solutions in $AdS_5 \times S^5$},
Nucl.\ Phys.\ B {\bf 668} (2003) 77;
{\tt hep-th/0304255}.}

\lref\Minahan{
J.~A.~Minahan,
{\it Higher loops beyond the SU(2) sector},
JHEP {\bf 0410}, 053 (2004);
{\tt hep-th/0405243}.
}

\lref\Engquist{
J.~Engquist,
{\it Higher conserved charges and integrability for spinning strings in
$AdS_5 \times S^5$},
JHEP {\bf 0404}, 002 (2004)
{\tt hep-th/0402092}.
}

\lref\BSint{
N.~Beisert and M.~Staudacher,
{\it The N = 4 SYM integrable super spin chain},
Nucl.\ Phys.\ B {\bf 670}, 439 (2003);
{\tt hep-th/0307042}.
}

\lref\Nikthesis{
N.~Beisert,
{\it The dilatation operator of N = 4 super Yang-Mills theory and
integrability}; {\tt hep-th/0407277}.
}

\lref\Niktwothree{
N.~Beisert,
{\it The su(2$|$3) dynamic spin chain},
Nucl.\ Phys.\ B {\bf 682}, 487 (2004);
{\tt hep-th/0310252}.
}

\lref\ROW{
E.~Ogievetsky, P.~Wiegmann and N.~Reshetikhin,
{\it The Principal Chiral Field In Two-Dimensions On Classical Lie
Algebras: The Bethe Ansatz Solution And Factorized Theory Of Scattering},
Nucl.\ Phys.\ B {\bf 280}, 45 (1987).
}

\lref\BKS{
N.~Beisert, V.~A.~Kazakov and K.~Sakai,
{\it Algebraic curve for the SO(6) sector of AdS/CFT};
{\tt hep-th/0410253}.}

\lref\KMMZ{
V.~A.~Kazakov, A.~Marshakov, J.~A.~Minahan and K.~Zarembo,
{\it Classical/quantum integrability in AdS/CFT},
JHEP {\bf 0405}, 024 (2004);
{\tt hep-th/0402207}.
}

\lref\KZ{
V.~A.~Kazakov and K.~Zarembo,
{\it Classical/quantum integrability in non-compact sector of AdS/CFT};
{\tt hep-th/0410105}.}

\lref\SS{
M.~Staudacher, {\it The Factorized S-Matrix of CFT/AdS};
{\tt hep-th/0412188}.
}

\lref\Saleur{
H.~Saleur,
{\it The continuum limit of sl(N/K) integrable super spin chains},
Nucl.\ Phys.\ B {\bf 578}, 552 (2000);
{\tt solv-int/9905007}.
}

\lref\Essler{
F.~H.~L.~Essler and V.~E.~Korepin,
{\it Spectrum of Low-Lying Excitations in a Supersymmetric Extended Hubbard Model};
{\tt cond-mat/9307019}.
}

\lref\Schoutens{
K.~Schoutens,
{\it Complete solution of a supersymmetric extended Hubbard model},
Nucl.\ Phys.\ B {\bf 413}, 675 (1994).
}

\lref\Schlottmann{
P.~Schlottmann, 
{\it Theormodynamics of the degenerate supersymmetric t-J model in one
dimension}; J. Phys.: Condens. Matter {\bf 4}, 7565 (1992).
}

\lref\Takahashi{
M.~Takahashi,
{\it Many-body Problem of Attractive Fermions with Arbitrary Spin in
One Dimension}; Prog. Theo. Phys. {44}, 899 (1970).
}

\lref\TakahashiReal{
M.~Takahashi,
{\it One-Dimensional Electron Gas with Delta-Function Interaction at
Finite Temperature}; Prog. Theo. Phys. {46}, 1388 (1971).
}

\lref\KuResh{
P.~P.~Kulish and N.~Y.~Reshetikhin,
{\it Diagonalization Of Gl(N) Invariant Transfer Matrices And Quantum N Wave
System (Lee Model)},
J.\ Phys.\ A {\bf 16}, L591 (1983).
}

\lref\Kulish{
P.~P.~Kulish,
{\it Integrable Graded Magnets},
J.\ Sov.\ Math.\  {\bf 35}, 2648 (1986)
[Zap.\ Nauchn.\ Semin.\  {\bf 145}, 140 (1985)].
}

\lref\bmn{
D.~Berenstein, J.~M.~Maldacena, H.~Nastase,
{\it Strings in flat space and pp-waves from N = 4 super Yang Mills},
JHEP {\bf 0204}, 013 (2002);
{\tt hep-th/0202021}.}

\lref\AF{
G.~Arutyunov and S.~Frolov,
{\it Integrable Hamiltonian for classical strings on $AdS_5 \times S^5$};
{\tt hep-th/0411089}.}

\lref\ReshWieg{
N.~Y.~Reshetikhin and P.~B.~Wiegmann,
{\it Towards The Classification Of Completely Integrable Quantum Field
Theories},
Phys.\ Lett.\ B {\bf 189}, 125 (1987).
}

\lref\thooft{
G.~'t Hooft,
{\it A Planar Diagram Theory For Strong Interactions},
Nucl.\ Phys.\ B {\bf 72}, 461 (1974).
}

\lref\maldacena{
J.~M.~Maldacena,
{\it The large N limit of superconformal field theories and supergravity},
Adv.\ Theor.\ Math.\ Phys.\  {\bf 2}, 231 (1998)
[Int.\ J.\ Theor.\ Phys.\  {\bf 38}, 1113 (1999)];
{\tt hep-th/9711200}
}

\lref\wittenads{E.~Witten,
{\it Anti-de Sitter space and holography},
Adv.\ Theor.\ Math.\ Phys.\  {\bf 2}, 253 (1998);
{\tt hep-th/9802150}.
}

\lref\GKP{
S.~S.~Gubser, I.~R.~Klebanov and A.~M.~Polyakov,
{\it A semi-classical limit of the gauge/string correspondence},
Nucl.\ Phys.\ B {\bf 636}, 99 (2002);
{\tt hep-th/0204051}.}

\lref\BDS{
N.~Beisert, V.~Dippel and M.~Staudacher,
{\it A novel long range spin chain and planar N = 4 super Yang-Mills},
JHEP {\bf 0407}, 075 (2004);
{\tt hep-th/0405001}.}

\lref\FT{
S.~Frolov and A.~A.~Tseytlin,
{\it Semiclassical quantization of rotating superstring in $AdS_5 \times S^5$},
JHEP {\bf 0206}, 007 (2002);
{\tt hep-th/0204226}.}

\lref\tseytlin{
A.~A.~Tseytlin,
{\it Semiclassical strings and AdS/CFT},
{\tt hep-th/0409296}.
}

\lref\Onish{
A.~V.~Kotikov, L.~N.~Lipatov, A.~I.~Onishchenko and V.~N.~Velizhanin,
{\it Three-loop universal anomalous dimension of the Wilson operators in N = 4
 SUSY Yang-Mills model},
Phys.\ Lett.\ B {\bf 595}, 521 (2004);
{\tt hep-th/0404092}.
}

\lref\MZ{
J.~A.~Minahan and K.~Zarembo,
{\it The Bethe-ansatz for N = 4 super Yang-Mills},
JHEP {\bf 0303}, 013 (2003);
{\tt hep-th/0212208}.
}

\lref\ocv{
J.~Lucietti, S.~Schafer-Nameki and A.~Sinha,
{\it On the exact open-closed vertex in plane-wave light-cone string
field theory},
Phys.\ Rev.\ D {\bf 69}, 086005 (2004);
{\tt hep-th/0311231}.
}

\lref\LP{
M.~Luscher and K.~Pohlmeyer,
{\it Scattering Of Massless Lumps And Nonlocal Charges In The
Two-Dimensional
Classical Nonlinear Sigma Model},
Nucl.\ Phys.\ B {\bf 137}, 46 (1978).}

\lref\Nikstrings{
N.~Beisert,
{\it Higher-loop integrability in N = 4 gauge theory},
Comptes Rendus Physique {\bf 5}, 1039 (2004);
{\tt hep-th/0409147}.}

\lref\BPR{
I.~Bena, J.~Polchinski and R.~Roiban,
{\it Hidden symmetries of the $AdS_5 \times S^5$ superstring},
Phys.\ Rev.\ D {\bf 69}, 046002 (2004);
{\tt hep-th/0305116}.}

\lref\DNW{
L.~Dolan, C.~R.~Nappi and E.~Witten,
{\it A relation between approaches to integrability in superconformal
Yang-Mills theory},
JHEP {\bf 0310}, 017 (2003)
{\tt hep-th/0308089}.}

\lref\DN{
L.~Dolan and C.~R.~Nappi,
{\it Spin models and superconformal Yang-Mills theory};
{\tt hep-th/0411020}.}

\lref\AFS{
G.~Arutyunov, S.~Frolov and M.~Staudacher,
{\it Bethe ansatz for quantum strings},
JHEP {\bf 0410}, 016 (2004)
{\tt hep-th/0406256}.}

\lref\AFRT{
G.~Arutyunov, S.~Frolov, J.~Russo and A.~A.~Tseytlin,
{\it Spinning strings in $AdS_5 \times S^5$ and integrable systems},
Nucl.\ Phys.\ B {\bf 671}, 3 (2003);
{\tt hep-th/0307191}.}

\lref\ART{
G.~Arutyunov, J.~Russo and A.~A.~Tseytlin,
{\it Spinning strings in $AdS_5 \times S^5$: New integrable system relations},
Phys.\ Rev.\ D {\bf 69}, 086009 (2004);
{\tt hep-th/0311004}.
}

\lref\BFST{
N.~Beisert, S.~Frolov, M.~Staudacher and A.~A.~Tseytlin,
{\it Precision spectroscopy of AdS/CFT},
JHEP {\bf 0310}, 037 (2003);
{\tt hep-th/0308117}.
}

\lref\BQS{
N.~Beisert,
{\it Spin chain for quantum strings};
{\tt hep-th/0409054}.}

\lref\ArutyunovStaudacher{
G.~Arutyunov and M.~Staudacher,
{\it Matching higher conserved charges for strings and spins},
JHEP {\bf 0403}, 004 (2004);
{\tt hep-th/0310182}.
}

\lref\ASII{
G.~Arutyunov and M.~Staudacher,
{\it Two-loop commuting charges and the string / gauge duality};
{\tt hep-th/0403077}.
}

\lref\SerbanS{
D.~Serban and M.~Staudacher,
{\it Planar N = 4 gauge theory and the Inozemtsev long range spin
chain},
JHEP {\bf 0406}, 001 (2004);
{\tt hep-th/0401057}.}

\lref\Lipatov{
A.~V.~Kotikov and L.~N.~Lipatov,
{\it DGLAP and BFKL equations in the N = 4 supersymmetric gauge
theory},
Nucl.\ Phys.\ B {\bf 661}, 19 (2003),
[Erratum-ibid.\ B {\bf 685}, 405 (2004)];
{\tt hep-ph/0208220}
}

\lref\Vbraun{
A.~V.~Belitsky, V.~M.~Braun, A.~S.~Gorsky and G.~P.~Korchemsky,
{\it Integrability in QCD and beyond},
{\tt hep-th/0407232}.
}

\lref\BraunII{
V.~M.~Braun, S.~E.~Derkachov and A.~N.~Manashov,
{\it Integrability of three-particle evolution equations in {QCD}},
Phys.\ Rev.\ Lett.\  {\bf 81}, 2020 (1998);
{\tt hep-ph/9805225}.
}

\lref\StringSpin{
N.~Beisert, J.~A.~Minahan, M.~Staudacher and K.~Zarembo,
{\it Stringing spins and spinning strings},
JHEP {\bf 0309}, 010 (2003);
{\tt hep-th/0306139}.
}


\Title{\vbox{
\hbox{hep-th/0412254}
\hbox{DESY-04-250}
}}
{\vbox{\centerline{The Algebraic Curve of 1-loop Planar ${\cal N}=4$ SYM}}}
\bigskip
\centerline{Sakura Sch\"afer-Nameki}
\bigskip
\centerline{\it II. Institut f\"ur Theoretische Physik der
Universit\"at Hamburg}
\centerline{\it Luruper Chaussee 149, 22761 Hamburg, Germany}
\centerline{\tt sakura.schafer-nameki@desy.de}
\vskip2.1cm
\centerline{\bf Abstract}
\bigskip
\noindent
The algebraic curve for the $\psu (2,2|4)$ quantum spin chain
is determined from the thermodynamic limit of the algebraic Bethe
ansatz. The Hamiltonian of this spin chain has been identified with
the planar 1-loop dilatation operator of ${\cal N}=4$ SYM. In the dual
$AdS_5\times S^5$ string theory, various properties of the data
defining the curve for the gauge theory are compared to
the ones obtained from semiclassical
spinning-string configurations, in particular for the case of 
strings on $AdS_5 \times S^1$ and the $\su(2,2)$ spin chain agreement of the curves is shown.

\bigskip

\Date{12/2004}


\newsec{Introduction}

Integrable structures in $d=4$ ${\cal N}=4$ $SU(N)$ Super-Yang Mills
theory (SYM) have recently been utilized to put to 
test gauge/string holography \refs{\thooft} realized in terms of the
AdS/CFT correspondence \refs{\maldacena, \wittenads}. 
Extending the seminal work of Minahan and Zarembo \refs{\MZ}, the key
observation of Beisert and Staudacher \refs{\BeisertI, \BSint} is the
identification of the 1-loop dilatation operator of planar ${\cal N}=4$ SYM with the Hamiltonian of a quantum spin chain for the
Lie super-algebra $\psu(2,2|4)$. Put into the context of the AdS/CFT
correspondence, one would expect to find a corresponding integrable
structure in the $AdS_5\times S^5$ string theory. Evidence to this
effect was obtained by 
following the proposals in \refs{\bmn, \GKP} by
Frolov and Tseytlin \refs{\FT, \Frolov} and subsequently in \refs{
\StringSpin, \AFRT,\BFST, \ART}\footnote{$^\sharp$}{For further
references on semi-classical spinning strings see \eg\
\refs{\tseytlin}.} by considering semi-classical string
configurations, with large spins on the $AdS_5$ and/or
$S^5$.\footnote{$^\natural$}{Related QCD-based caculations have
appeared in \refs{\Lipatov, \Onish}. 
For references on integrability
in QCD see \refs{\BraunII, \Vbraun}.} 
Another line of research successfully matched the
local charges of the integrable systems \refs{\ASII, \Engquist}, and comparisons
of the non-local charges have appeared in
\refs{\BPR, \DNW}. Progress towards a quantum Bethe-ansatz for the
(notoriously
difficult to quantize) $AdS_5\times S^5$ string was made in
\refs{\ArutyunovStaudacher, \AFS, \BQS}.
Integrability seems to persist beyond 1-loop
\refs{\SerbanS, \ASII, \BDS, \Minahan}, however a mismatch has emerged at
3-loops, the origin of which has been conjectured to be an order of limits
problem \refs{\BDS, \Nikthesis}\footnote{$^\flat$}{A similar order
of limits issue is discussed in \refs{\ocv}.}. 

The main objective of this paper is to expand on the methods of
\refs{\KMMZ, \KZ, \BKS}, the idea of which is the following:
An integrable system can be characterized by an algebraic
curve, which is constructed out of the transfer matrix, and in particular contains information about the local
charges. In the case of quantum spin chains the curve is extracted from the
transfer matrix in the thermodynamic limit, which in the context of
AdS/CFT would need to be
compared to the string sigma-model curve for large spins in the limit
$\lambda/J^2\rightarrow 0$, where $\sqrt{\lambda}$ is the string tension and
$J$ the angular momentum on the $S^5$. 
One would expect that a necessary condition 
for the integrity of the AdS/CFT correspondence is the agreement of
the algebraic curves of the respective quantum/classical integrable
systems. As a first step, in this note the curve
for the full ${\cal N}=4$ SYM theory will be constructed. 

The agreement of the gauge and string theory curves in the $\su(2)$ and $\slfrak(2)$ subsectors
was proven in \refs{\KMMZ, \KZ} and was then shown to hold in
the $\so(6)$ subsector of AdS/CFT by Beisert, Kazakov
and Sakai \refs{\BKS}.
In this note these methods are applied in a straight forward manner to the
$\su(2,2)$ subsector, which is closed at 1-loop
order \refs{\BeisertI} and the resulting curve is identified with the one for the sigma-model
on $AdS_5\times S^1$. 

The present note is structured as follows: in section 2 the super-spin
chain for $\psu(2,2|4)$ is discussed, including a recap of the Bethe
ansatz of \refs{\BSint} and a derivation of the transfer matrix
eigenvalues. Following a succinct discussion of the termodynamic limit
in section 3, the algebraic curve is derived from the Bethe
equations in section 4. A comparison to some known properties of the full
$AdS_5\times S^5$ sigma-model curve is given in section 5 and the
agreement of the curves in the $\su(2,2)$ subsector and the $AdS_5
\times S^1$ sigma-model, respectively, is shown in section 6. We conclude in section
7. Appendix A collects formulae for transfer-matrices of super-spin
   chains and in appendix B the algebraic curve for the $\su(2,2)$ spin chain with
   anti-ferromagnetic ground state, which is of relevance for the QCD
   spin chain \refs{\BFHZ}. 


\newsec{The $\psu(2,2|4)$ Spin Chain}

In this section some known facts about spin chains with
Lie super-algebra symmetries are reviewed, flashing out various points that are
different from the more common case of Lie
algebras. Apart from fixing notation and spelling out the Bethe equations, we
give an explicit expression for the eigenvalue of the transfer matrix
for the $\psu(2,2|4)$ spin chain, which will be essential for
constructing the algebraic curve. 

\subsec{R-matrix}

The Lie super-algebra of interest for the spin chain associated to the
dilatation operator of ${\cal N}=4$ SYM is $\psu(2,2|4)$. More
precisely, the planar 1-loop dilatation operator of
this theory has been shown to be identical to the Hamiltonian of a
$\psu(2,2|4)$ integrable spin chain \refs{\BeisertI, \BSint}. The
integrability was infered by an explicit construction of an R-matrix. 
Denote by $\R_{ij}(u)$ the R-matrix acting on the tensor product
$\V_i\otimes \V_j$, where $\V_i$ denotes the $\psu(2,2|4)$-module
located at site $i$ of the spin chain and $\V_0=\V_{aux}$ be the auxiliary space. The
R-matrix satisfies the Yang-Baxter equations
\eqn\YB{
\R_{ij}(u_i-u_j) \R_{ik}(u_i-u_k) \R_{jk}(u_j-u_k) =
\R_{jk}(u_j-u_k) \R_{ik}(u_i-u_k) \R_{ij}(u_i-u_j) \,.
}
Further define the transfer matrix along the entire spin chain of
length $L$ by 
\eqn\transfermatdef{
\T (u) = \R_{01}(u)\R_{02}(u) \cdots \R_{0L}(u) \,.
}
It acts on the auxiliary space, where it will be formally written in block-form
\eqn\transferaux{
\T(u)=\pmatrix{A(u) & B(u)\cr C(u)& D(u)} \,.
}
The simplest choice for $\V_{aux}$ is the fundamental representation,
\ie, the $\fourfour$ for $\psu(2,2|4)$, for the trace of which we
shall provide an expression below.

\subsec{Bethe equations}

For a Lie algebra or Lie super-algebra $\g$, 
define a set of Bethe roots $u_i^{(k)}$, $k=1,\cdots, r=\rank (\g)$ and
$i=1,\cdots, J_k$, where $J_k$ denotes the excitation
number for the $k$th root.  
Further, define $J=\sum J_k$ as the total excitation number and let
$L$ be the length of the
spin chain. Then the corresponding Bethe equations were
determined in \refs{\ROW} to be
\eqn\gbethe{
\left({u_i^{(k)} - {i\over 2} V_{k}\over u_i^{(k)} + {i\over 2} V_{k}} \right)^L
= \prod_{l=1}^{r}\prod_{j=1}^{J_l}  {u_i^{(k)} - u_j^{(l)}
-{i\over 2}M_{k l}\over  u_i^{(k)} - u_j^{(l)} +{i\over 2}M_{k l}}\,.
}
Translational invariance along the spin chain implies further that
\eqn\betheextra{
1 = \prod_{k=1}^r 
\prod_{i=1}^{J_k} {u_i^{(k)} + {i\over 2} V_{k} \over u_i^{(k)} -
{i\over 2} V_{k}} = e^{ i P} \,.
}
The $\g$-dependent data entering these equations are the 
Cartan matrix $M_{k l}$
and the vector of Dynkin labels $V_{k}$ of the representation that is located
at the respective spin chain sites. Eq. \betheextra\ yields
furthermore a quantization condition for the total momentum $P$. 

The validity of this Bethe ansatz (most of the Bethe ansatz
methodology goes thought for super-algebras, up to some sign changes, which are
detailed in appendix A) for Lie
super-algebras has been established in \refs{\Kulish}. 
The Cartan matrix for $\psu(2,2|4)$ which is most suitable for the present analysis has been
discussed in \refs{\BSint, \Nikthesis} and is based on the Dynkin
diagram in Figure 1. 

\fig{`Beauty' Dynkin diagram for $\psu(2,2|4)$.}{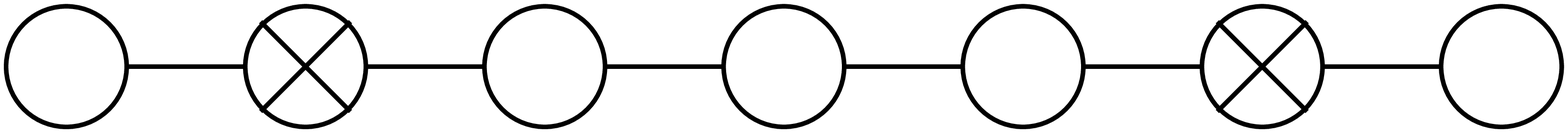}{1.8truein}

\noindent
The nodes $\otimes$ denote as usual the fermionic roots. 
The choice of Dynkin diagram for Lie super-algebras is not unique, and
an alternative choice is discussed in the appendix, which will be
useful for the reduction to $\su(2,2)$ in section 6. 
The Cartan matrix for the Dynkin diagram in figure 1 can be put into the form 
\eqn\Cartanmat{
M=
\pmatrix{
-2& 1 &    & & & &  \cr
1 & 0 & -1 & & & &  \cr
  & -1 & 2 &-1 & & &\cr
  &  & -1 & 2& -1& & \cr
  &  & &-1 &2 &-1 & \cr
  &  & & &-1 & 0 &1  \cr
  &  & & & & 1 & -2  
}\,.
}
In fact, this is the Cartan matrix only up to rearrangements of rows,
which however leave the Bethe equations invariant, \cf\ \refs{\Nikthesis}.

In the present setup, single-trace operators in the SYM theory 
are identified with 
states of a spin chain with periodic boundary conditions, where the
spins at each site transform in a representation of
$\psu(2,2|4)$. The Dynkin labels of the highest weights will be
denoted by
\eqn\hwt{
V= (s_1, r_1; q_1, q, q_2; r_2, s_2) \,,
}
where $[q_1, q, q_2]$ are the Dynkin labels for $\so(6)$ and $[s_1,
s_2]$ of $\so(3,1)$. Each state is characterized further by
the bare dimension $\Delta_0$, the length $L$ of the spin-chain as
well as the hyper-charge $B$.
The
multiplet, in which the elementary component fields of ${\cal N}=4$
SYM transform, and which will thus be the representation at the
lattice sites of the spin chain, is the so-called field-strength
multiplet and has highest weight vector
\eqn\frep{
V_{\bf F} = (0,0;0,1,0;0,0) \,.
}

\subsec{Transfer-matrix Eigenvalues}

From now on we fix the representation at site $i$ to be $\V_i={\bf
F}$ and consider various choices for $\V_{aux}$. 
The super-trace of the transfer matrix in the auxiliary space 
\eqn\Tmatevaldef{
T_{\bf R}(u) = \STr_{\bf R} \T(u)\,,\qquad \V_{aux}={\bf R} \,,
}
contains the integrals of motion, or local/global charges, when
expanded in the spectral parameter $u$. 
Let us choose $\V_{aux}= \CCop^N=\CCop^{4+4}$ to be the fundamental representation
and consider the decomposition of $\T(u)$ in
\transferaux\ into blocks acting on $\CCop^{N-1}\oplus \CCop$, so that $C_k(u)$ is an $(N-1)$-column.
The state space is then generated by these oscillators $C_{k}(u_i^{(k)})$ 
\eqn\states{
|\{ u_i^{(k)}\} , L\rangle = \prod_{k=1}^r \prod_{i=1}^{J_k}
C_k\left(u_i^{(k)}\right) |0, L\rangle\,.
}
The action of $\T(u)$ upon it is determined from the exchange
relations between $A$ and $D$ with $C$ resulting from the Yang-Baxter equations \YB. 
Define
\eqn\Rfu{
R_k(u) = \prod_{i=1}^{J_k} (u-u_i^{(k)}) \,.
}
The eigenvalue of the trace of the transfer matrix for the $\ufrak(4|4)$ spin chain is computed for $\V_{aux}= \fourfour$ in
appendix A, and for the SYM spin-chain it is
\eqn\Tmateval{\eqalign{
{(u+i/2)^L\over u^L } T_{\fourfour }(u)
& ={(u+i/2)^L\over u^L}\, {R_1(u+i)\over R_1(u)}  \cr
& +{(u+i/2)^L\over u^L}\, {R_1(u-i)\over R_1(u)} {R_2(u+i)\over R_2(u)}   \cr
& -{(u+i/2)^L\over u^L}\, {R_2(u+i)\over R_2(u)} {R_3(u-i)\over R_3(u)}   
  -{(u+i/2)^L\over u^L}\, {R_3(u+i)\over R_3(u)} {R_4(u-i)\over R_4(u)}   \cr
& -{(u-i/2)^L\over u^L}\, {R_4(u+i)\over R_4(u)} {R_5(u-i)\over R_5(u)}   
  -{(u-i/2)^L\over u^L}\, {R_5(u+i)\over R_5(u)} {R_6(u-i)\over R_6(u)}   \cr
& +{(u-i/2)^L\over u^L}\, {R_6(u-i)\over R_6(u)} {R_7(u+i)\over R_7(u)}   \cr
& +{(u-i/2)^L\over (u+i)^L}\quad\, {R_7(u-i)\over R_7(u)} \,.
}}
The local charges of the spin chain are extracted from the trace of
the transfer matrix with $\V_{aux}= \V_{\bf F}$. The direct
computation of this is involved, as the representation in question is
infinite-dimensional. However, the relation between
$T_{\bf F}$ and the local charges $Q_r$ is
\eqn\TQ{
T_{\bf F} (u+u_0) = \exp \left(i \sum_{r=2}^\infty  u^{r-1} Q_r \right) \,,
}
where the expansion of the transfer matrix trace is around the point $u_0$,
where the R-matrix reduces to a projector. The coefficients $Q_r$ in
this expansion give rise to the local charges. 
This expansion is an important input for the construction of the
algebraic curve and has been quoted in \refs{\ReshWieg, \Nikthesis} 
\eqn\Tasymp{
T_{\bf R} (u+u_0) = \prod_{l=1}^r \prod_{j=1}^{J_r} 
{u-u_j^{(l)} - {i\over 2} V_l \over u-u_j^{(l)} + {i\over 2} V_l} +
O(u^L) \,.
}
Note that this is precisely the term that appears in the cyclicity
constraint \betheextra. 
For ${\bf R}= {\bf F}$ the asymptotics are
\eqn\TasympFrep{
T_{\bf F} (u+u_0) = \prod_{j=1}^{J_4} 
{u-u_j^{(4)} - {i/2} \over u-u_j^{(4)} + {i/ 2} } + O(u^L) 
= {R_4(u-{i\over 2}) \over R_4 (u+ {i\over 2})} + O(u^L) \,.
}
Further, the asymptotics of the higher local charges are
\eqn\Qasymp{
Q_r = {i\over r-1} \sum_{l=1}^r \sum_{j=1}^{J_r} 
\left(
{1\over (u_j^{(l)} + {i\over 2} V_l)^{r-1}} - 
{1\over (u_j^{(l)} - {i\over 2} V_l)^{r-1}}  
\right)\,.
}


\newsec{Thermodynamic Limit}

The first step in determining the algebraic curve of the above system
is to consider the thermodynamic limit $L\rightarrow \infty$. This
procedure has been explained for the $\su(2)$ spin chain in
\refs{\KMMZ} and will now be discussed for general $\g$. 
Taking the logarithm of \gbethe\ yields
\eqn\logbethe{
L \log \left({u_i^{(k)} - {i\over 2} V_{k}\over u_i^{(k)} + {i\over 2} V_{k}} \right)
= \sum_{l=1}^r \sum_{j=1,\, j\not= i}^{J_l} \log \left(
{u_i^{(k)} - u_j^{(l)} -{i\over 2}M_{k l}
\over  u_i^{(k)} - u_j^{(l)} +{i\over 2}M_{k l}}
\right) - 2\pi i n_i^{(k)}\,,
}
where $n_i^{(k)}\in \Zop$ are the mode numbers, arising due to taking the logarithm. 
Define $x_i^{(k)} = u_i^{(k)}/L$ and perform the large $L$ and $J$ limit of this equation
\eqn\largeL{
 {V_{k}\over x_i^{(k)}} = \sum_{l=1}^r {1\over J_l}\sum_{j=1,\,  j\not=
i}^{J_l} {M_{k l}\over
x_i^{(k)}-x_j^{(l)}} - 2\pi n_i^{(k)} \,.
}
For fixed $n_k$ define the densities
\eqn\densities{
\rho_k(x) = 
\sum_{j=1}^{J_k} \delta (x-x_j^{(k)}) \,,
}
and the corresponding resolvents
\eqn\resolvent{
G_k(x) = {1\over J_k} \sum_{j=1}^{J_k} {1 \over x-x_j^{(k)}} \,.
}
In the limit, the following replacement is made
\eqn\thermolimit{
\sum_{j=1}^{J_k} \rightarrow L \int_{\C_k} dv \rho_k(v) \,,
}
where ${\cal C}_k$ denotes the curve along which the Bethe roots condense. The densities are normalized as
\eqn\densnorm{
\int_{\C_k} \rho_k(u) = J_k \,.
}
The resolvent become
\eqn\resolventt{
G_k(u)= {1\over J_k}\int_{{\cal C}_k} dv \, {\rho_k(v)  \over v-u} \,,
}
and the Bethe equations limit to
\eqn\bethedense{
\slashint_{\C} dv \, {\rho_k(v) M_{k f(v)}\over v-u} = -{V_{k}\over u} + 2\pi n_i^{(k)} \,, \qquad u\in \C_i^{(k)}\,,
}
where $\C=\cup_k \C_k$ and each of the curves $\C_k$ associated to
simple roots is on the other hand $\C_k=\cup_j \C_j^{(k)}$. Further, 
$f(u)= l$ for $u\in\C_l\subset\C$.
The principal value $\Slashint$ results from the restriction of the sum in
\largeL\ to $j\not=i$.
Equivalently we have
\eqn\betheresolved{
M_{k k} \Gslash_k(u)+  \sum_{l\not= k}  M_{k l} G_l(u) =
-{V_{k} \over u} + 2\pi n_i^{(k)}  \,,\qquad u \in \C_i^{(k)} \,. 
}
Slashes denote principal values, that is 
\eqn\slashdef{
\Gslash (u)=\half ( G(u+) + G(u-))\,.
}

For the discussion of the algebraic curve and its properties we shall
require the asymptotics of the resolvents around $u=\infty$. The
analysis is identical to the one in \refs{\BKS}. At infinity their
expansion is
\eqn\Gasymp{
G_k(u) = -{1\over u} \int_{\C_k} dv \rho_k(v) + O\left({1\over u^2}\right) 
= -{J_k \over u} + O\left({1\over u^2}\right) \,.
}
The relation between the excitations numbers and the Dynkin labels
\hwt\ was
obtained in \refs{\BSint}
\eqn\extodyn{
J_k= \pmatrix{
&\half \Delta_0 -\half (L-B) - \half s_1 \cr
 &     \Delta_0 -       (L-B)  \cr
  &    \Delta_0 -\half       (L-B)  -\half q - {3\over 4} q_1
      -{1\over 4} q_2 \cr
&\Delta_0  - q -\half q_1 -\half q_2 \cr
&\Delta_0 -\half(L+B)  -\half  q -{1\over 4} q_1 -{3\over 4} q_2 \cr
&\Delta_0 - (L+B) \cr
&\half \Delta_0 -\half (L+B) - \half s_2
} \,.
}


\newsec{The Algebraic Curve}

The algebraic curve is implicitly defined in the Bethe equations
\betheresolved. In the standard procedure to relate \betheresolved\ to an
algebraic curve a set of quasi-momenta
$p^{\bf R}= (p_1(u),\cdots, p_8(u))$ is defined. One way to determine them
directly is to compute the transfer matrices for $\psu(2,2|4)$ in the
representation ${\bf R}$ and extract the $p_i$ from their large $L$
limit. For the $\so(6)$ sector this was done in \refs{\BKS}. Here the
same procedure is applied to the transfer matrix eigenvalue \Tmateval. 


\subsec{Fermionic roots}

In the Lie super-algebra case there is an additional subtlety due to
the fermionic Lie algebra roots: as these square to zero, the corresponding
diagonal entry in the Cartan matrix vanishes, which reflects itself
in the absence of the diagonal, self-interacting term for the Bethe roots
associated to the fermionic Lie algebra root. The above argument for
deriving the thermodynamic limit needs to be treated with some care,
as the
corresponding continuum equation ceases to be a singular integral
equation.

We shall discuss the case of the fermionic root $\alpha_2$, the case
of $\alpha_6$ works in an identical fashion. 
The first four Bethe equations, which are relevant for this discussion
with $V= V_{\bf F}$
are
\eqn\FBeq{\eqalign{
1&= \prod_{j\not= i}^{J_1}  {u_i^{(1)} - u_j^{(1)} +i
\over  u_i^{(1)} - u_j^{(1)} -i} \ 
\prod_{j=1}^{J_2}  {u_i^{(1)} - u_j^{(2)}- {i\over 2}
\over  u_i^{(1)} - u_j^{(2)} +{i\over 2}} \cr
1&= \prod_{j=1}^{J_1}  {u_i^{(2)} - u_j^{(1)}
-{i\over 2}\over  u_i^{(2)} - u_j^{(1)} +{i\over 2}} \qquad\qquad\qquad\qquad
\prod_{j=1}^{J_3}  {u_i^{(2)} - u_j^{(3)}
+{i\over 2}\over  u_i^{(2)} - u_j^{(3)} -{i\over 2}} \cr
1&= \qquad\qquad\qquad\qquad 
\prod_{j=1}^{J_2}  {u_i^{(3)} - u_j^{(2)} +{i\over 2}\over  u_i^{(3)} - u_j^{(2)} -{i\over 2}}
\prod_{j\not= i}^{J_3}  {u_i^{(3)} - u_j^{(3)} -i
\over  u_i^{(3)} - u_j^{(3)} +i} \
\prod_{j=1}^{J_4}  {u_i^{(3)} - u_j^{(4)}-{i\over 2}\over  u_i^{(3)} - u_j^{(4)} +{i\over 2}}\cr
V(u_i^{(4)})^L
&= \qquad\qquad\qquad\qquad\qquad\qquad\qquad\qquad\ 
\prod_{j=1}^{J_3}  {u_i^{(4)} - u_j^{(3)}
+{i\over 2}\over  u_i^{(4)} - u_j^{(3)} -{i\over 2}}
\prod_{j\not= i}^{J_4}  {u_i^{(4)} - u_j^{(4)} -i
\over  u_i^{(4)} - u_j^{(4)} +i} 
\,,
}}
where $V(u_i^{(4)})$ is the RHS of \gbethe.
The roots $u^{(k)}_i=0$ for $k=5,6,7$, which is admissible.
First note that the excitation numbers satisfy the bounds $0\leq J_1
\leq J_2 \leq J_3 \leq J_4$. Secondly, the bosonic Bethe roots come in complex conjugate pairs, unless they are real. 

Consider the simpler case when in addition $u^{(1)}_i=0$. The second equation
in \FBeq\ turns into
\eqn\realitycon{
1= \prod_{j=1}^{J_3}  {u_i^{(2)} - u_j^{(3)} +{i\over 2}\over
u_i^{(2)} - u_j^{(3)} -{i\over 2}} \,,
}
which is an algebraic equation of degree $J_3-1$ in $u^{(2)}_i$. For
each $u^{(3)}_k$,  $\bar{u}^{(3)}_k$ appears as well. From the second
equation follows then by taking the absolute value that $u^{(2)}_i\in
\Rop$ (a related argument appears in \refs{\TakahashiReal}).  
So in this case, the solutions to the Bethe equations are of the type
that all bosonic Bethe roots come in complex conjugate pairs (or are real)
and the fermionic Bethe roots are real and solutions to the algebraic
equation \realitycon. 

Including the $u^{(1)}_i$ roots, 
the second equation yields a polynomial equation of degree $J_1 +
J_3-1$ for the Bethe roots $u_i^{(2)}$. Reality does not follow in
this case. 
The trick to solve this was introduced in the
condensed matter literature in 
\refs{\Takahashi} and used further in \refs{\Essler, \Schoutens,
\Saleur}. The idea is to introduce a string of Bethe roots for each
fermionic Bethe root. In order to spell out the procedure for the
present purposes, make the ansatz for the fermionic roots $u_k^{(2)}$ and
$u_k^{(6)}$ as follows\footnote{$^\ast$}{A proof, that this
string-ansatz yields the complete set of solutions, does not seem to
exist.}
\eqn\FermiBetheRoots{\eqalign{
u_k^{(2)} &  
=\left\{ \eqalign{
v_k^{(2)}       &\qquad k=1,\cdots, J_2 \cr
v_k^{(1)} + i/2 &\qquad k=J_2 +1 ,\cdots, J_2+J'_1  \cr
v_k^{(1)} - i/2 &\qquad k=J'_1 +J_2 + 1 ,\cdots, 2 J'_1+J_2
          }
\right. \cr
u_k^{(6)} &
=\left\{ \eqalign{
v_k^{(6)}       &\qquad k=1,\cdots, J_6 \cr
v_k^{(7)} + i/2 &\qquad k=J_6 +1 ,\cdots, J_6+J'_7  \cr
v_k^{(7)} - i/2 &\qquad k=J_6 +J'_7 + 1 ,\cdots, J_6+2J'_7\,,
          }
\right. 
}}
and values $v^{(1)}_k$ and $v^{(7)}_k$ for $u^{(1)}_k$ and
$u^{(7)}_k$, respectively, 
where all $v_j^{(k)} \in \Rop$. This has to hold strictly only in the
$L=\infty$ limit. The total number of Bethe roots for $\alpha_2$ is
then $J_2 + 2 J_1'$ {\it etc.}. 

We shall discuss the case of the fermionic root $\alpha_2$, the case
of $\alpha_6$ works in an identical fashion. 
Plugging \FermiBetheRoots\ into the second set of Bethe equations gives
\eqn\FBeq{\eqalign{
1&= \prod_{j=1}^{J_1'}  {v_i^{(1)} - u_j^{(1)} -i
                        \over  v_i^{(1)} - u_j^{(1)} +i} \
    \prod_{j\not= i }^{J_2}  {v_i^{(1)} - v_j^{(1)}- i
                        \over  v_i^{(1)} - v_j^{(1)} + i} \
    \prod_{j=1}^{J_3}  {v_i^{(1)} - u_j^{(3)} +i
                       \over  v_i^{(1)} - u_j^{(3)} -i} \cr
1&= 
\prod_{j=1}^{J_1'}  {v_i^{(2)} - u_j^{(1)} -{i\over 2}
                        \over  v_i^{(2)} - u_j^{(1)} +{i\over 2}} 
    \prod_{j=1 }^{J_2}  {v_i^{(2)} - v_j^{(1)}- {i\over 2}
                        \over  v_i^{(2)} - v_j^{(1)} + {i\over 2}} 
    \prod_{j=1}^{J_3}  {v_i^{(2)} - u_j^{(3)} +{i \over 2}
                       \over  v_i^{(2)} - u_j^{(3)} -{i\over 2}}
\,.
}}
In particular, the Bethe roots $v_i^{(1)}$ now have a self-interacting
term, as opposed to the real Bethe roots $v_i^{(2)}$ associated to
$\alpha_2$. Switching the roots $v_i^{(1)}$ off reduces the Bethe
equation to one without self-interactions, for the real roots $v_i^{(2)}$. 

So in summary the distribution of Bethe roots comprises of complex
Bethe roots assigned to the bosonic Lie algebra roots, which condense
in complex contours (denoted ${\cal C}_i^{(k)}$), as well as real
Bethe roots associated to the fermionic Lie algebra roots, and
triplets of roots (\eg, for $\alpha_2$ these are $v_k^{(2)},
v_k^{(2)}\pm i/2$), assigned to $\alpha_{1/2}$ and $\alpha_{6/7}$. 
The real roots associated to the fermionic roots are algebraically
determined in terms of the other Bethe roots.


\subsec{Construction of the Algebraic Curve}

For the present case of interest let us first spell out the Bethe
equations \betheresolved\ for the bosonic Lie algebra roots
\eqn\bethedetail{\eqalign{
-2 \Gslash_1 (u) + G_2(u) 
& 
=2\pi n_j^{(1)} - {V_{j_1} \over u}\,, \qquad u\in \C_j^{(1)} \cr
-  G_2 (u) +2 \Gslash_3(u) - G_4 (u) 
&
= 2\pi n_j^{(3)}- {V_{j_3} \over u}\,, \qquad u\in \C_j^{(3)} \cr
-  G_3 (u) +2 \Gslash_4(u) - G_5 (u) 
&
= 2\pi n_j^{(4)}- {V_{j_4} \over u}\,, \qquad u\in \C_j^{(4)} \cr
-  G_4 (u) +2 \Gslash_5(u) - G_6 (u) 
&
= 2\pi n_j^{(5)}- {V_{j_5} \over u}\,, \qquad u\in \C_j^{(5)} \cr
G_6 (u) -2 \Gslash_7(u) 
&
= 2\pi n_j^{(7)} - {V_{j_7} \over u}\,, \qquad u\in \C_j^{(7)}\,. 
}}
Each of the above lines with mode number $n_j^{(k)}$ corresponds to
the $k$th root and it is assumed that $u\in {\cal C}_{j}^{(k)}$. {\it
I.e.}, for each root the densities $\rho_k(u)$ have support on the union of curves
$\C_k= \C_1^{(k)} \cup\cdots \cup \C_{A_k}^{(k)}$.
The Bethe equations associated to the fermionic roots give rise to the
algebraic constraints
\eqn\fermionicBethe{\eqalign{
{1\over J_1} \sum_{j=1}^{J_1} {1\over u^{(2)}_i - u^{(1)}_j}
- {1\over J_3} \sum_{j=1}^{J_3} {1\over u^{(2)}_i - u^{(3)}_j}
&= 2\pi n_i^{(2)} \cr
{1\over J_6} \sum_{j=1}^{J_5} {1\over u^{(6)}_i - u^{(5)}_j}
- {1\over J_7} \sum_{j=1}^{J_7} {1\over u^{(6)}_i - u^{(7)}_j}
&= 2\pi n_i^{(6)}\,.
}}
In the case of only real values for the fermionic 
Bethe roots, $u^{(k)}_j = v_j^{(k)}$ for $k=2,6$, these can be rewritten as
\eqn\RealFermi{\eqalign{
G_1 (u) - G_3(u)
&
= 2\pi n_j^{(2)}- {V_{j_2} \over u}\,, \qquad u\in {\cal R}^{(2)} \cr
-G_5 (u) + G_7(u)
&
= 2\pi n_j^{(2)}- {V_{j_2} \over u}\,, \qquad u\in {\cal R}^{(6)} \,,
}}
where ${\cal R}^{(k)}\subset \Rop$, so \RealFermi\ are
algebraic equations, which have to be satisfied for a collection of
real points. 

The situation changes, if we allow complex
values for the fermionic roots. Then, the thermodynamic limit needs to
be taken for the equations \FBeq. Let $H_k(u)$ be the resolvent for
the additional real centers $v_j^{(k)}$, $k=1,7$, thus
\eqn\ComplexFermi{\eqalign{
G_1(u) + \Hslash_1(u) - G_3(u) &=  \pi m_j^{(2)} \,, \qquad  u\in {\cal R}^{(2)} \cr
G_1(u) + H_1(u) -G_3(u) &= 2 \pi l_j^{(2)} \,, \qquad\,  u\in {\cal S}^{(2)} \cr
G_7(u) + \Hslash_7(u) - G_5(u) &=  \pi m_j^{(6)} \,, \qquad  u\in {\cal R}^{(6)} \cr
G_7(u) + H_7(u) -G_5(u) &= 2 \pi l_j^{(6)} \,, \qquad\,  u\in {\cal S}^{(6)} 
\,.
}}
The principal value now
arises from the self-interacting term for the Bethe root $v_k^{(1)}$
in \FBeq. The second equation is again algebraic and both
equations are for real values of $u$.

Now we turn to the construction of the algebraic curve. Assume that the fermionic Bethe roots are real.  
The aim is to rewrite the
equations \bethedetail\ in 
terms of the quasi-momenta $p_k(u)$, which will be defined shortly, such
that they take the form 
\eqn\quasibethe{
M_{kk}\tilde{\Gslash}_{k} +\sum_{j\not=k} M_{kj} \Gt_j(u) = \pslash_k(u) -\pslash_{k+1}(u) = 2 \pi n_j^{(k)}\,,
 \qquad u\in \C_j^{(k)} \,,
}
$k\not= 2,6$, 
where the singular terms in \bethedetail\ have been absorbed into the
 resolvents, $\Gt_k(u)$. For the fermionic roots, the Bethe equations
 in terms of quasi-momenta are
\eqn\FermiQuasiBethe{
\sum_{j\not=k} M_{kj} \Gt_j(u) = p_k(u) - p_{k+1}(u) = 2 \pi
 n_j^{(k)}\,,\qquad k=2,6\,,
}
for $u\in {\cal R}_j^{(k)}$ real. 

\fig{First four sheets of the algebraic curve.}{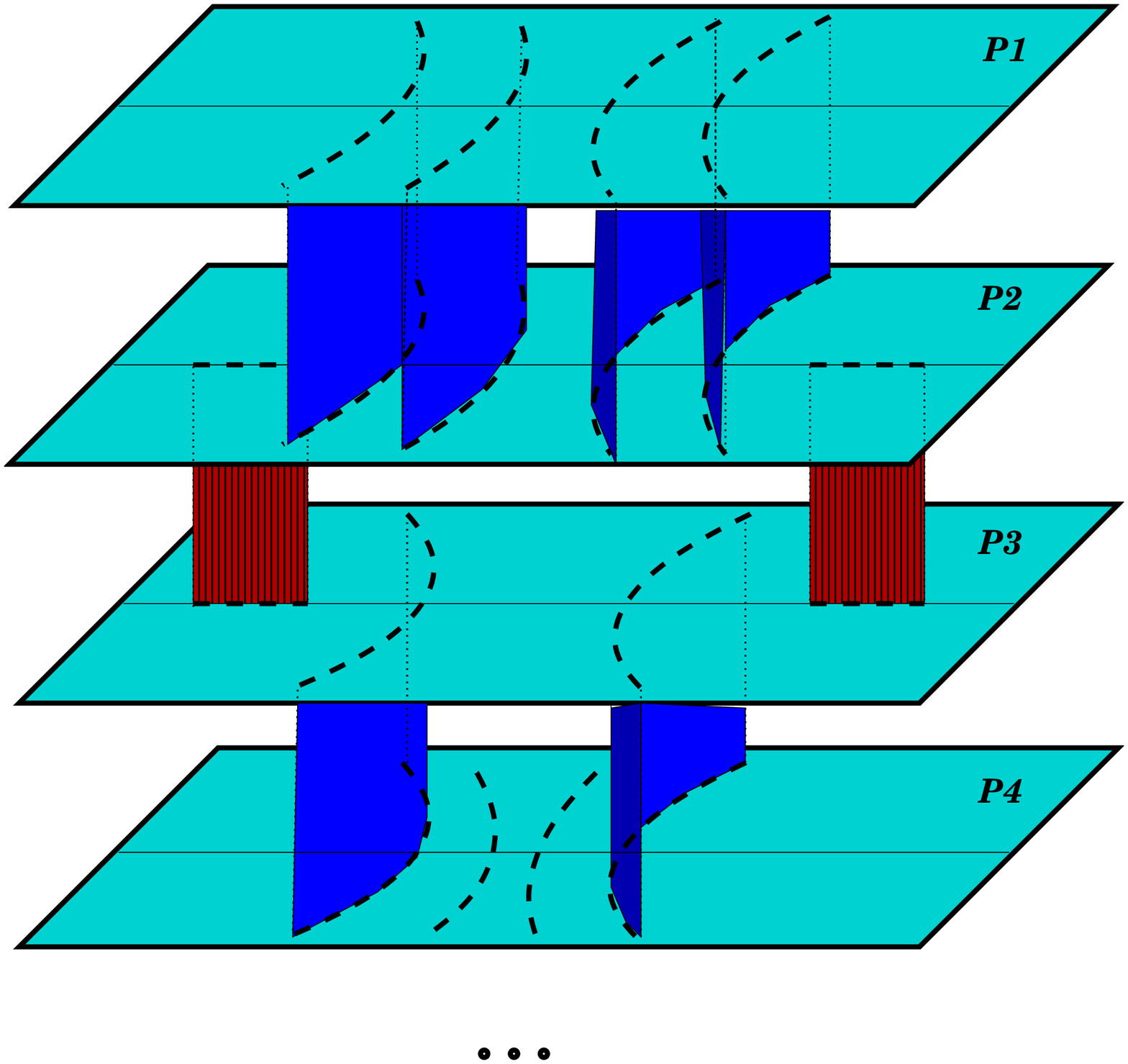}{2.4truein}

\noindent
In summary: the various sheets of the algebraic curve are labeled by
$p_k(u)$, and for bosonic roots are glued together along the cuts ${\cal
C}_k$, whereas for the fermionic roots the Bethe roots lie on
the real axis, where the quasi-momenta of the sheets satisfy an
algebraic relation. In figure 2 the first four sheets assigned to
$p_k(u)$ are depicted schematically. The curves within each sheet are Bethe roots distributed along
the complex cuts ${\cal C}^{(k)}_j$ for the bosonic roots, and the real Bethe roots on sheets
$2$ and $3$ lie on the real axis. The vertical dotted lines indicate the
identifications of the sheet functions, encoded in \quasibethe\ (blue) and
\FermiQuasiBethe\ (red stripy).

The quasi-momenta $p_k$ are obtained from the eigenvalue $T_{\fourfour}$ of the transfer
matrix \Tmateval\ in the large $L$ limit 
\eqn\Tlimit{
T_{\fourfour}(u) \rightarrow \sum_{k=1}^8 \epsilon_k \exp(i p_k) \,, 
}
with (\cf \Tmateval)
\eqn\eps{
\epsilon_k= (++----++)\,. 
} 
The terms in \Tmateval\ limit to
\eqn\Rlimit{
{R_k(u+s) \over R_k(u+t)} \rightarrow \exp ((t-s) G_k(u)) \,,\qquad
{(u+s)^L\over (u+t)^L }\rightarrow \exp \left({(s-t)\over u}\right) \,.
}
The representation relevant for the SYM spin chain has Dynkin labels
$V_{\bf F}$, see \frep. Define the singular resolvents by
\eqn\singG{\eqalign{
\Gt_1(u) &= G_1(u) - {1\over 2u} \,,\qquad
\Gt_2(u)  = G_2(u) - {1\over u}  \,,\qquad
\Gt_3(u)  = G_3(u) - {1\over 2u} \,,  \cr
\Gt_4(u) &= G_4(u)   \,, \cr
\Gt_5(u) &= G_5(u) - {1\over 2u}   \,,\qquad
\Gt_6(u)  = G_6(u) - {1\over u}   \,,\qquad
\Gt_7(u)  = G_7(u) - {1\over 2u}   \,.
}}
The quasi-momenta for this representation then turn out to be
\eqn\pdef{\eqalign{
p_1(u) &= -\Gt_1(u)         \cr
p_2(u) &= -\Gt_2(u) +\Gt_1(u)  \cr
p_3(u) &= +\Gt_3(u) -\Gt_2(u)  \cr
p_4(u) &= +\Gt_4(u) -\Gt_3(u)  \cr
p_5(u) &= +\Gt_5(u) -\Gt_4(u)  \cr
p_6(u) &= +\Gt_6(u) -\Gt_5(u)  \cr
p_7(u) &= -\Gt_7(u) +\Gt_6(u)  \cr
p_8(u) &= +\Gt_7(u) \,,
}}
and need to satisfy the Bethe equations \bethedetail\ in their
incarnation \quasibethe. 
Note that
\eqn\psum{
\sum_{k=1}^8  \epsilon_k p_k =0 \,.
}
With these redefinitions, the Bethe equations take the compact form of
a Riemann-Hilbert problem
\eqn\bethefinal{
\pslash_k(u) -\pslash_{k+1}(u) = 2\pi n_j^{(k)} \,,\qquad u\in\C_j^{(k)} \,,
}
for $k=1,3,4,5,7$. In addition, the algebraic equations
\fermionicBethe\ have to be satisfied. 
The functions $p_k(u)$ with $k=1,\cdots,8$ in \bethefinal\ determine a
function $p_{\bf F}(u)$ defined on a Riemann surface, where each $p_k$ is
$p_{\bf F}(u)$ restricted to the $k$th sheet. 

The asymptotics of the quasi-momenta at $u=\infty$ are obtained from
\Gasymp\ and \extodyn 
\eqn\quasiasymp{\eqalign{
p_1(u) &=  {1\over u} \left(+ \half \Delta_0 -\half s_1  \right) + O\left({1\over u^2}\right) \cr 
p_2(u) &=  {1\over u} \left(+ \half\Delta_0+ \half s_1  \right) + O\left({1\over u^2}\right)\cr 
p_3(u) &=  {1\over u} \left(+{3\over 4} q_1+\half q + {1\over 4} q_2 \right)+ O\left({1\over u^2}\right) \cr 
p_4(u) &=  {1\over u} \left(-{1\over 4} q_1+\half q + {1\over 4} q_2 \right)+ O\left({1\over u^2}\right)  \cr 
p_5(u) &=  {1\over u} \left(-{1\over 4} q_1-\half q + {1\over 4} q_2 \right)+ O\left({1\over u^2}\right) \cr 
p_6(u) &=  {1\over u} \left(-{1\over 4} q_1-\half q - {3\over 4} q_2 \right)+ O\left({1\over u^2}\right)\cr 
p_7(u) &=  {1\over u} \left(- \half\Delta_0- \half s_2 \right)+ O\left({1\over u^2}\right) \cr
p_8(u) &=  {1\over u} \left(- \half\Delta_0+ \half s_2 \right)+ O\left({1\over u^2}\right)\,.
}}
Thus, all quasi-momenta have the asymptotics $p_i(u)\sim 1/u+ O(1/u^2)$.

From the relation between the quasi-momenta and resolvents the
asymptotics at $u=0$ are
\eqn\pasympzero{
p_k(u)= \left\{
\eqalign{
+{1\over 2u}+ O(1) & \qquad k=1,2,3,4\cr
-{1\over 2u}+ O(1) & \qquad k=5,6,7,8 \,.
}
\right.
}
The asymptotics of the local
charge expansion \Qasymp\ is obtained as follows, taking into account that the expansion
around $u_0$ becomes an expansion around $u=0$ in the thermodynamic
limit. Eq. \TasympFrep\ in the thermodynamic limit yields by
\Rlimit\footnote{$^\diamondsuit$}{There is a slightly subtle point here: in
\TasympFrep\ only the terms up to order $u^L$ are determined. However,
as explained in sec. 4.5 of \refs{\Nikthesis}, these terms may well
contribute in the thermodynamic limit, yielding terms of
$O(1/u)$. However assuming that the term will be a combination of
quasi-momenta $p_k$, 
the $O(1/u)$ term can be determined indirectly from the asymtotics
of $p_k(u)$ at $u=0$.}
\eqn\thermoTG{
\Gt_4(u) = G_4(u)= p_1(u) + p_2(u) + p_3(u) + p_4(u) =
{2\over u} + \sum_{r=1}^\infty u^{r} Q_r \,. 
}
In summary, by \quasiasymp\ and \pasympzero, the function $p(u)$ defined by the quasi-momenta on each of the eight
sheets is therefore not regular. An algebraic curve can be engineered
out of it by removing the singularities. Consider 
\eqn\curveparameter{
y(u) = \epsilon u^2 {dp(u)\over du} \,.
}
or written on each of the sheets
\eqn\ysheetwise{
y_k(u) = \epsilon_k u^2 {dp_k(u)\over du} \,.
}
$y(u)$ then has no poles and satisfies an octic equation 
\eqn\octic{
\sum_{d=0, d\not= 7}^8 P_d(u)y^d = P_8(u) \, \prod_{d=1}^8 (y-y_k(u)) =0 \,,
}
where $y_k(u) = u^2 p_k'(u)$ and the coefficient of $y^7$ vanishes by \psum. The argument runs essentially the same
way as in \refs{\BKS}: $p(u)$ has poles at $0$ and $\infty$. Assume that
$y(u)$ takes constant values at these points. Then \octic\ is
satified only if the $P_k(u)$
have the same degree $2 d$ and the constant term does not vanish --
in this case \octic\ gives rise to a polynomial equation in $y$ with
constant coefficients, which can be solved for finite $y$. The curve
thus has $8(2 d +1)-1$ parameters left.  
As explained in \refs{\BKS} it is necessary to ensure the right number
of square-root branch cuts, which are determined from the square-root
poles in $y$. These are $d$ in number. 
The absence of other unwanted cuts is obtained by requiring the discriminant for the polynomial on the LHS of \octic\ 
\eqn\discriminant{
D(y) = P_8(u)^{14} \prod_{i<j}^8 (y_i -y_j)^2 \,,
}
to be a perfect square.
The discriminant condition reduces the number of parameters by $13
d$. Further, the pole-structure at $u=0$ fixes further $7$
coefficients, leaving $3d$ parameters, which are fixed by the
period-integrals: $d$ parameters are fixed by requiring that the
A-cycle period-integrals (\ie, integrals circumnavigating one cut within
a sheet) vanish. Further, the Bethe equations in the
Riemann-Hilbert form are equivalent to the integrality of the
B-periods (\ie, integrals along cycles, which connect two sheets) with the integers given by the $n_j^{(k)}$ in the Bethe
equations. This fixes another $d$ parameters. Finally, choosing the filling fractions gives rise
to the remaining $d$ parameters, of which one is fixed by the momentum
condition resulting from \betheextra, \cf\
\refs{\KMMZ}.\footnote{$^\ddagger$}{Thanks to the authors of
\refs{\BKS} for making ${\tt v3}$ of their paper available in advance,
where this is discussed in the $SO(6)$ case.}


\newsec{Comments on the Algebraic Curve for the Sigma-model on $AdS_5\times S^5$}

Some very crude comparisons between the ${\cal N}=4$ SYM algebraic curve
and the (yet-to-be-determined) algebraic curve for the string theory
on $AdS_5\times S^5$ can be made. In the spirit of the discussions in
\refs{\KMMZ, \KZ, \BKS} the idea is that the properties of the
quasi-momenta determine various properties of the curve, as \eg, seen
in the last section. Thus, a necessary
requirement for the matching of the two curves is that the asymptotics
of the $p_k$ agree. 

On the string theory side, some important properties of the
quasi-momenta were determined recently by Arutyunov and Frolov
\refs{\AF} by constructing a Lax representation of the full string
sigma-model. We shall compare the there-obtained asymptotics of the
quasi-momenta with the ones derived in section 4.  

\subsec{Asymptotics of the Sigma-Model Quasi-Momenta}

Denote the spectral parameter of the string sigma-model by $x$. Then
the asymptotics for the quasi-momenta were obtained in \refs{\AF} for
$x$, which will be compared to the asymptotics in the spectral
parameter $u$ used so far in this paper.

The expressions for $p_k(u)$ for
$u=\infty$ obtained in \quasiasymp\ are mapped to the ones in \refs{\AF}
by identifying the representation labels in \extodyn\
with the various spins of
$AdS_5$ ($S_1, S_2$) and $S^5$ ($J_1, J_2, J_3$) by means of 
\eqn\afrel{\eqalign{
q_1 &= J_2-J_3 \,,\, \qquad  q_2 = J_1 -J_2 \,,\qquad  q = J_2+ J_3
\,,\cr
s_1 &= S_1 - S_2 \,,\qquad s_2 = S_1 + S_2 \,.
}}
By \refs{\AF}, the asymptotics of the quasi-momenta $q_i(x)$ of the sigma-model at
$x=0$ and $x=\infty$ are\footnote{$^\star$}{An overall minus sign is introduced compared
to \refs{\AF}, which does not spoil the symmetry of
$p_k(x)$ under $x\rightarrow 1/x$ and simplifies the comparison to the
gauge theory.}
\eqn\sigmapasymp{ 
\eqalign{
(x&=0) \cr
q_1(x) &= x {2\pi \over \sqrt{\lambda}} (-\Delta_0+ S_1 - S_2) \cr
q_2(x) &= x {2\pi \over \sqrt{\lambda}} (-\Delta_0- S_1 + S_2)\cr 
q_3(x) &= x {2\pi \over \sqrt{\lambda}} (+J_3 -J_1  -J_2)\cr
q_4(x) &= x {2\pi \over \sqrt{\lambda}} (-J_3 -J_1  +J_2)\cr 
q_5(x) &= x {2\pi \over \sqrt{\lambda}} (-J_3 +J_1  -J_2)\cr
q_6(x) &= x {2\pi \over \sqrt{\lambda}} (+J_3 +J_1  +J_2)\cr 
q_7(x) &= x {2\pi \over \sqrt{\lambda}} (+\Delta_0+ S_1 + S_2)\cr
q_8(x) &= x {2\pi \over \sqrt{\lambda}} (+\Delta_0- S_1 - S_2)\,,
}   
\qquad 
\eqalign{
(x& =\infty) \cr
q_1(x) &= {2\pi \over x \sqrt{\lambda}} (+\Delta_0 - S_1 + S_2) \cr
q_2(x) &= {2\pi \over x \sqrt{\lambda}} (+\Delta_0 + S_1 - S_2)\cr 
q_3(x) &= {2\pi \over x \sqrt{\lambda}} (-J_3 +J_1 +J_2)\cr
q_4(x) &= {2\pi \over x \sqrt{\lambda}} (+J_3 +J_1 -J_2)\cr 
q_5(x) &= {2\pi \over x \sqrt{\lambda}} (+J_3 -J_1 +J_2)\cr
q_6(x) &= {2\pi \over x \sqrt{\lambda}} (-J_3 -J_1 -J_2)\cr 
q_7(x) &= {2\pi \over x \sqrt{\lambda}} (-\Delta_0- S_1 - S_2)\cr
q_8(x) &= {2\pi \over x \sqrt{\lambda}} (-\Delta_0+ S_1 + S_2)\,.
}
}
The quasi-momenta $q_k(x)$ in \sigmapasymp\ for $k=3,4,5,6$
reproduce the ones for the five-sphere, and for $k=1,2,7,8$ for $AdS$.
The asymptotics at $x\rightarrow \pm 1$ are vastly more complicated
((5.5) and (5.8) in \refs{\AF}).

\subsec{Comparison to the $\psu(2,2|4)$ Spin Chain}

The comparison between gauge theory and string theory is made by
taking the limit
$L\rightarrow \infty$ in the SYM theory (which has been accounted for by considering the
spin chain in the thermodynamic limit) and likewise the
large angular momentum limit of the string theory. More precisely, define $J=\sum J_i$ and
$S= \sum S_i$, and consider the limit
\eqn\limits{
{\lambda\over J^2} \rightarrow 0\,,\qquad 
J \rightarrow \infty \,,\qquad 
{J_i\over J}, \ {S_i\over J} =\hbox{fixed} <\infty \,.
}
First, the relation between the spectral parameters
$u$ and $x$ needs to be sorted out. 
From the asymptotics of the quasi-momenta at $\infty$ one
naively reads off ${1\over u}\sim {4\pi\over x \sqrt{\lambda}}$. In order
to match the $x, u\rightarrow 0$ asymptotics, one infers that 
\eqn\specpara{
u= {\sqrt{\lambda} \over 4 \pi J} x \,.
}
Rescaling the various spins ($S_i, \Delta_0, J_i$) by $J$, the
asymptotics at $u,x=\infty$, \quasiasymp\ and \sigmapasymp, agree.

At $u=0$ the comparison is a bit more subtle. From the gauge theory
result \pasympzero\ we expect simple poles with strengths $\pm 1/2$ at
$u=0$. Setting $u=\pm {\sqrt{\lambda} \over 4 \pi J}$, which amounts
to $u\rightarrow 0$ in the limit \limits, implies by \specpara\ that
this is to be compared with the limit $x\rightarrow \pm 1$ of the
string theory quasi-momenta, which are rather involved expressions,
whose $\lambda/J^2\rightarrow 0$ limit is not easily extraced. 
We shall discuss this elsewhere in more detail. 


Clearly it would be desirable to construct the complete curve for the sigma-model on
$AdS_5\times S^5$, and perform a comparison of the curves, as is done
in the case of $\Rop\times S^5$ in \refs{\BKS}
and for $AdS_5 \times S^1$ in section 6.


\newsec{The Algebraic Curve for the $\su(2,2)$ `Subsector'}

In this section the reduction of the algebraic 
curve to the subsector which is the dual to spinning strings on
$AdS_5\times S^1$ is considered. The curve and Bethe ansatz are
obtained for the sigma-model and are
shown to agree with the ones of the $\su(2,2)$ spin chain. A word of caution is in place here: we should point out that the trunction to
$\su(2,2)$ does not yield a subsector of ${\cal N}=4$ in
general, however it is a closed subsector at 1-loop \refs{\BeisertI}. One can truncate the spin chain to
this symmetry algebra and show that the quasi-momenta have the same
asymptotics with the
ones obtained for the $AdS_5\times S^1$ sigma-model. The corresponding
spinning string solutions with multiple $AdS$-spins and one spin on
the $S^5$ were first obtained in \refs{\Frolov}.

\subsec{The $\su(2,2)$ Spin Chain}

The reduction from $\psu(2,2|4)$ to $\su(2,2)$ is most 
transparent by picking the distinguished Dynkin diagram of $\psu(2,2|4)$,
as done in appendix A\footnote{$^\flat$}{Recently the $\su(2,2)$ spin
chain has also been discussed in the context of integrability in large
N QCD \refs{\BFHZ}. 
There, the representation at each site is ${\bf
F}$ 
.
}. The quatities computed with this choice will be
distinguished with a prime. The first question to address is which
$\su(2,2)$-representation is to sit at each spin chain
site. The sub-module of ${\bf F}$, which is obtained by acting with
$\su(2,2)$ on the highest weight $V_{\bf F}$ (which for the
distinguished Dynkin diagram is $V_{{\bf F}'}=(0, -3, 2)$) would be one
choice. However the highest weight state here corresponds in the SYM
theory to a highly excited state, and thus discussing a termodynamic
limit in this case does not make sense.  
Nevertheless the curve in this case is computed in appendix B, as it
is of relevance for the QCD spin chain \refs{\BFHZ}.  

The correct choice for the representation for the $\su(2,2)$ subsector
of ${\cal N}=4$ SYM has highest weight
$V'_{\bf Z}=(0,-1,0)$. It is built on the physical
vacuum and thus the Bethe ansatz equations describe excitations around
this BPS-state, whereas ${\bf F}'$ would have a vacuum energy of $3L$ \refs{\BSint}
and the excitations would be built on a non-BPS state and thus unstable\footnote{$^\sharp$}{Thanks to N. Beisert for explaining
this point to me.}. 
This yields quasi-momenta, with
asymptotics matching the ones of the string sigma-model.
The non-compact $\su(2,2)$-module with highest weight $(0,-1,0)$ 
corresponds in ${\cal N}=4$ SYM to
the sector with vacuum $|0\rangle = \Tr Z^L$, with the scalar
$Z=\Phi_{34}$, which in the oscillator representation of \refs{\BeisertI}
is $|Z\rangle = {\bf c}_3^\dagger {\bf c}_4^\dagger |0\rangle$. The states are
obtained by acting on $|0\rangle$ with the oscillators ${\bf
a}^\dagger_i{\bf b}^\dagger_{j}$, $i,j\in\{1,2\}$, which correspond to
space-time derivatives ${\cal D}_{ij}$. 

In order to determine the asymptotics of the quasi-momenta, one
requires to first find the
relation between the representation labels and excitation numbers. In
\refs{\BeisertI} the excitation numbers $n_{\bf a}$ {\it etc.} for the oscillators were
obtained for a state of the type 
$({\bf a}^\dagger_1)^{k_1} ({\bf a}^\dagger_2)^{k_2} ({\bf
b}^\dagger_1)^{l_1} ({\bf b}^\dagger_2)^{k_1+k_2 +l_1}
|0\rangle$. Furthermore, the combination of oscillators corresponding
to the roots $\alpha_k$ of $\su(2,2)$ are 
\eqn\roottoosci{
\alpha_1\,:\quad {\bf b}_2^\dagger {\bf b}^1\,,\qquad
\alpha_2\,:\quad {\bf b}_1^\dagger {\bf a}_1^\dagger\,,\qquad
\alpha_3\,:\quad {\bf a}_2^\dagger {\bf a}^1\,,
}
so that the excitation numbers $J'_k$ for the roots are
$J_1=n_{{\bf b}_2}$, $J_2= n_{{\bf b}_2}+ n_{{\bf b}_1}$ and $J_3=
n_{{\bf b}_2}+ n_{{\bf b}_1} -n_{{\bf a}_1}= n_{{\bf a}_2}$.
In summary, the relation between excitation numbers for the roots and
the representation labels are thus
\eqn\suttrel{
J'_k = \pmatrix{
& -\half + &\half (\Delta_0 -s_2) \cr
& -1 + &\Delta_0 \cr
& -\half  + &\half (\Delta_0 -s_1) 
}\,.
}

The singular resolvents and quasi-momenta are extracted from the
transfer matrix in appendix A, where all the $G'_k=0$ for $k=4\cdots
7$
\eqn\Gprime{\eqalign{
\Gt'_1 (u) &= G'_1(u) - {1\over 2 u} \cr
\Gt'_2 (u) & = G'_2 (u) - {1\over u} \cr
\Gt'_3(u) & = G'_3(u) - {1\over 2 u}\,,
}}
and
\eqn\pprime{\eqalign{
p'_1(u) &= \Gt'_1(u) \cr
p'_2(u) &= \Gt'_2(u) - \Gt'_1(u) \cr
p'_3(u) &= \Gt'_3(u) - \Gt'_2(u) \cr
p'_4(u) & = -\Gt'_3(u) \,.
}}
The corresponding Bethe equations in the large $L$ limit take the form
as expected from the general considerations in \betheresolved
\eqn\betheprime{\eqalign{
2 \Gslash'_1 (u) - G'_2(u) 
&=  \pslash'_1 (u) - \pslash'_2 (u) - {V'_{j_1} \over u}
=2\pi n_j^{(1)} - {V'_{j_1} \over
u}\,, \qquad  u\in \C_j^{(1)} \cr
- G'_1(u) + 2 \Gslash'_2 (u) - G'_3(u) 
&=  \pslash'_2 (u) - \pslash'_3 (u) - {V'_{j_2} \over u}
=2\pi n_j^{(2)} - {V'_{j_2} \over
u}\,, \qquad  u\in \C_j^{(2)} \cr
2 \Gslash'_3 (u) - G'_2(u) 
&=  \pslash'_3 (u) - \pslash'_4 (u) - {V'_{j_3} \over u}
=2\pi n_j^{(1)} - {V'_{j_3} \over
u}\,, \qquad  u\in \C_j^{(3)}  \,,
}}
where with the definition of the singular resolvents \Gprime, the RHS
is precisely given by $V'_{\bf Z}=(0,-1,0)$.  

The asymptotics of the quasi-momenta are extracted in the same way as
before from  \Gasymp\ and \suttrel\ at $u=\infty$
\eqn\pprimeasymp{\eqalign{
p'_1(u) &=  {1\over u} \left(- \half \Delta_0 +\half s_2  \right) + O\left({1\over u^2}\right) \cr 
p'_2(u) &=  {1\over u} \left(- \half\Delta_0  -\half s_2  \right) + O\left({1\over u^2}\right)\cr 
p'_3(u) &=  {1\over u} \left(+ \half \Delta_0 +\half s_1  \right) + O\left({1\over u^2}\right) \cr 
p'_4(u) &=  {1\over u} \left(+ \half\Delta_0  -\half s_1  \right) +
O\left({1\over u^2}\right) \,.
}}
The asymptotics at $u=0$ are obtained from \Tasymp. At finite $L$ the
expansion is
\eqn\Tprimeasymp{
T_{\bf Z} (u+i) = 
{R_2(u+i/2 )\over R_2(u-i/2)}   + O(u^L)\,,
}
which for $L\rightarrow \infty$ limits to 
\eqn\Tprimelim{
T_{\bf Z} (u+i) \rightarrow \exp \left(-i \Gt'_2(u))\right) \,.
}
Thus at $u=0$ the combination of resolvents 
\eqn\localcharges{
-\Gt'_2(u) =
-p_1'(u) - p_2'(u) =
 {1\over  u} +
\sum_{r=0}^\infty u^r Q_r \,. } 
gives rise to the local
charges $Q_r$. The asymptotics of all quasi-momenta at zero are
\eqn\pprimezero{
p'_1(u) = -{1\over 2u}+ O(1)\,,\quad 
p'_2(u) = -{1\over 2u}+ O(1)\,,\quad
p'_3(u) = +{1\over 2u}+ O(1)\,,\quad
p'_4(u) = +{1\over 2u}+ O(1)\,.
}

The curve for this subsector is written in terms of $y_k = u^2 {d^2
p_k\over du^2}$ as
\eqn\curveprime{
\sum_{i=0, i\not= 3}^4 P_i(u) y^i = P_4 (u) \prod_{k=1}^4 (y- y_k(u)) =0 \,.
}
Again the moduli-count is as before and pretty much identical to the
$\so(6)$ case discussed in \refs{\BKS}: $\sum_k p'_k =0$, thus the $P_d$
together have $4 (2d +1) -1$ parameters, of which $3$ are fixed by
requiring the asymptotics at $u=0$. The discriminant condition
fixes $5d$ and the vanishing of the A-periods removes another $d$,
leaving $2d$, which are fixed by the B-period integrals ($d$), the
filling fractions ($d-1$) and momentum constraint ($1$). The latter
follows from integrating up the Bethe equations using that $G_2(u)$ at
order $u^0$ is $2\pi m$  
and reduces the number of $d$ filling fractions to $d-1$.

\subsec{Classical Sigma-model on $AdS_5\times S^1$}

Classically the string moving on $AdS_5\times S^1$ is described by a
coset model on 
\eqn\adscoset{
AdS_5 \times S^1 = {SO(4,2)\over SO(4,1)} \times U(1) \,.
}
The construction that will be most suited for a discussion of the integrable
structure is based on the Lax-pair representation obtained recently by
Arutyunov and Frolov in \refs{\AF}. 

For the comparison we quote the relevant properties of the sigma-model
integrable system in \refs{\AF}. 
Denote the spectral parameter on the sigma-model side by $x$. Then the
asymptotics of the quasi-momenta at $x=\infty$ are
\eqn\asympp{\eqalign{
q_1(x) &= {1\over x} {2\pi \over \sqrt{\lambda}} (-\Delta_0 +S_1 +S_2)\cr
q_2(x) &= {1\over x} {2\pi \over \sqrt{\lambda}} (-\Delta_0 -S_1 -S_2)\cr
q_3(x) &= {1\over x} {2\pi \over \sqrt{\lambda}} (+\Delta_0 +S_1 -S_2)\cr
q_4(x) &= {1\over x} {2\pi \over \sqrt{\lambda}} (+\Delta_0 -S_1 +S_2)\,,
}}
and at $x=0$ (there are no
non-trivial cycles and thus no windings
in the $AdS$ part)
\eqn\asymppzero{\eqalign{
q_1(x) &= x{2\pi \over \sqrt{\lambda}}  (+\Delta_0  -S_1 -S_2) \cr
q_2(x) &= x{2\pi \over \sqrt{\lambda}}  (+\Delta_0  +S_1 +S_2) \cr
q_3(x) &= x{2\pi \over \sqrt{\lambda}}  (-\Delta_0  -S_1 +S_2) \cr
q_4(x) &= x{2\pi \over \sqrt{\lambda}}  (-\Delta_0  +S_1 -S_2) \,,
}}
as well as $x=\pm 1$, with $m$ the winding number of the string on the
$S^1$,
\eqn\asymppmone{
q_1(x)= q_2(x)= -q_3(x)=-q_4 (x) = -{\pi \over x\mp 1} \left({J\over
\sqrt{\lambda} } \pm m\right) \,.
}
The spins are related to the ones appearing in the gauge theory
expressions by \afrel. 

The Lax connections formed out of left and right invariant currents
for the sigma-model are related by a gauge
tranformation and by replacing $x\rightarrow 1/x$ \refs{\AF}. The
associated transfer matrices are therefore related by the same
transformation, and the eigenvalues must agree up to the replacement
$x\rightarrow 1/x$
\eqn\xoneoverx{
q_i(x) = - q_i(1/x)  \,.
}
With these asymptotics of the quasi-momenta, the curve is constructed
as follows: First remove again the singularities, which now occur at
$x=0$ as well as $x=\pm 1$, and define the variables $z_k$ which are
regular on each of the four sheets defined by the quasi-momenta
$q_k$: $
z_k= x \left(x-{1\over x}\right)^2 {d q_k(x)\over dx}
$,
so that $z_k(1/x)=  z_k(x)$. $z$ satisfies a quartic equation of the
type \curveprime. The moduli count runs in an identical fashion to the
one for the $\so(6)$ curve in \refs{\BKS} and we refrain from
repeating it here and point out some differences:
the symmetry
\xoneoverx\ gives rise to the distribution of cuts such that the cut $\C_i$
connects the sheets $i$ and $i+1$. However as one readily verifies,
the moduli count is unchanged by this. Futher, the main difference is
the asymptotics of $q_k$ at $x=\pm 1$, which will in particular 
enter the Bethe equations. 
A point worth elaborating upon is the
identification of the
additional condition on the $d$ filling fractions. In the gauge
theory, the constraint arose from the momentum quantization condition,
which reduced the choice of filling fractions to $d-1$. It was derived
by integrating the Bethe equations in the Riemann-Hilbert form. In the
next subsection we shall derive the sigma-model analog of these.

\subsec{Bethe Equations for the Sigma-Model}

Using the isomorphism between $\su(2,2)$ and $\so(4,2)$, the
construction of the Bethe-type equations for the sigma-models proceeds
along the lines of $\so(6)$. Applying the methods put forward in \refs{\BKS}
and the asymptotics at $x=\pm 1$ in \asymppmone\ the resolvents
satisfy\footnote{$^\dagger$}{These equations are the analogs of (5.21)
in \refs{\BKS}, where the non-singular part of the resolvent has been
merged into $G_k$ and the part containing the singular term at $x=\pm
1$ is explicitly written out on the RHS.}
\eqn\Gsigma{\eqalign{
2 {\Gslash}_1 (x) - G_2 (x) &= 2\pi n_{j}^{(1)}
-  V_{1} {\frak L} \,,\qquad x \in \C_j^{(1)} \cr
-G_1(x)+ 2 {\Gslash}_2 (x) - G_3 (x) &= 2\pi n_{j}^{(2)} 
-  V_{2} {\frak L} \,, \qquad  x \in \C_j^{(2)} \cr
2 {\Gslash}_3 (x) - G_2 (x) &= 2 \pi n_{j}^{(3)} 
-  V_{3} {\frak L}
\,,\qquad  x \in \C_j^{(3)} \,,
}
}
where $V=(0, -1, 0)$ is again the vector of Dynkin labels for the
representation and 
\eqn\Ldef{
{\frak L} =
2 \pi \left({ {J\over \sqrt{\lambda}} + m \over x-1 } + { {J\over
\sqrt{\lambda}}- m \over x+1 }   \right) \,.
}
The equations \Gsigma\ are manufactured such that they are equivalent to
$\qslash_i -\qslash_{i+1} = 2\pi n_{j}^{(i)}$. 
The analog of the momentum constraint is then obtained by integrating the equations with
$\int_{{\cal C}_k} \rho_k(x)\cdot $ (where $\rho_k(x)$ are the
densities for the singular resolvents $\Gt_k(x)$) and using the normalization, $\int_{{\cal C}_k} \rho_k(x)/x
= -\Gt_k(0)$, which can be computed from the asymptotics of the
quasi-momenta at $x=0$ and \Ldef.

\subsec{Comparison Gauge/String Theory}

To show that the algebraic curves agree, the defining equations need
to be matched, as well as the asymptotics of the quasi-momenta need to
agree. The comparison is made at small $\lambda$ as before.
The spectral parameters $u$ and $x$ are related by
\eqn\spectralpara{
x= {4\pi J \over \sqrt{\lambda }} u\,.
}
Firstly, the number of sheets defining the curves, which is determined by
the number of quasi-momenta, agrees. Secondly, the asymptotics at
$u=\infty$, \pprimeasymp\ and \asympp, match after rescaling
$\Delta_0, S_1, S_2$ by $J$.

The asymptotics at $u=0$ are compared as follows. On the gauge theory
side, the asymptotics were determined in \pprimezero. The comparison
to the gauge theory requires to take the sigma-model quantities in the limit
\eqn\comppar{
{\lambda \over J^2} \rightarrow 0\,,\qquad 
\Delta_0 = S +J\,, \qquad J\rightarrow \infty\,,\qquad
{S_i\over J}=\alpha_i <\infty\,.
}
The asymptotics for the quasi-momenta at $u=0$ are obtained by setting
$u= \pm {\sqrt{\lambda}\over 4\pi J}$ (which in the limit \comppar\
becomes $u\rightarrow 0$) \ie, by considering the
asymptotics \asymppmone\ at $x=\pm 1$ in the sigma-model. Then
\eqn\qcalc{\eqalign{
\pm q_k(u\rightarrow 0) &= -{\pi \over x-1} \left({J\over
\sqrt{\lambda}+m}\right) - -{\pi \over x+1} \left({J\over
\sqrt{\lambda}-m}\right) \cr
&= -{1\over 4} \left({1\over u-{\sqrt{\lambda}\over 4 \pi J}}\left(1+
{\sqrt{\lambda} \over J} m\right)+    
{1\over u-{\sqrt{\lambda}\over 4 \pi J}}\left(1+
{\sqrt{\lambda} \over J} m\right)
\right) \,.
}}
So that in the limit $\lambda/J^2 \rightarrow 0$ we have
\eqn\qzero{
q_1(u) = -{1\over 2 u}\,,\quad 
q_2(u) =  -{1\over 2 u}\,,\quad
q_3(u) = {1\over 2 u}\,,\quad
q_4(u) = {1\over 2 u}\,.
}
The combination entering the generating function of local charges in
\localcharges\ has asymptotics
\eqn\complocalQ{
-q_1(u) -q_2(u)= {1\over u} + O(u^0)\,.
}
again setting $u= \pm {\sqrt{\lambda} \over 4 \pi J} \rightarrow 0$,
which is in agreement with the gauge theory asymptotics in \localcharges.
Together with the fact
that the defining equations for $y_k$ and $z_k$ are mapped into
eachother, this implies the agreements of the curves in the $\su(2,2)$
subsector. 

As an aside, note, that the comparison at $u=0$ breaks down for the $\su(2,2)$ spin
chain with representation ${\bf F}'$. The latter asymptotics were
computed in appendix B and do not yield the correct pole strengths
compared to \qzero. 
 
Finally, the Bethe equations \betheprime\ and \Gsigma, are shown to agree. This
follows by rewriting \Gsigma\ in terms of $u$ and then expanding the
RHS, \ie
\eqn\Bethecheck{\eqalign{
{\frak L} =& { {J\over \sqrt{\lambda}} + m \over x-1 } + { {J\over
\sqrt{\lambda}} - m \over x+1 } = {1\over 2 \pi } 
{
  u+{m\lambda \over 4\pi J^2 } \over 
u^2-{\lambda \over 16 \pi^2 J^2}
}\cr
=& {1\over 2 \pi} \left({1\over u} + \left[{1\over 16 \pi^2 u^3} +
{m\over 4\pi u^2}  \right] {\lambda \over J^2}+ O\left({\lambda^2\over
J^4}\right)\right) \,.
}}

So to lowest order $\lambda/J^2$ this reproduces the Bethe equations
obtained in the gauge theory. At order $\lambda/J^2$ an $m$-dependent
term appears, which via the Virasoro constraint can be seen to be a
non-local term. The corresponding two-loop result in the gauge theory is 
not known and it may in fact not be consistent to discuss the
$\su(2,2)$ sector at more than 1-loop. Nevertheless, truncating to 
$\slfrak(2)$, we get agreement with the result of \refs{\KZ}, where it was
suggested that the $m$-dependent term may be foreboding a breakdown
of the correspondence at two-loops. Very recently however it was shown
by Staudacher in \refs{\SS}, that in the $\sl(2)$ sector the agreement
goes through up to two-loops, the main point being that as proposed in
\refs{\BDS}, the map
between the spectral parameters
$u$ and $x$ is modified when including higher loop effects.


\newsec{Conclusions}

The main two objectives of this note were to adapt
the methods in \refs{\BKS} in order to construct the algebraic curve for the 
$\psu(2,2|4)$ super-spin chain as well as to
show the matching of the curve of the $\su(2,2)$ subsector and the
sigma-model for spinning strings on $AdS_5\times S^1$, respectively. The latter
analysis was a rather straight-forward application of the results on
the $\su(4)$ subsector, similar to the resemblence in the discussions
for the $\su(2)$ and $\slfrak(2)$ cases. The main difference being the
distinct asymptotics at $x=\pm 1$, which resulted in a prediction for
higher loops. A comparison to the gauge theory should be preceded by
an analysis along the lines of \refs{\Minahan}, checking, whether 
in the thermodynamic limit 
the $SU(2,2)$ sector is closed at higher loops.  

In appendix B an alternative algebraic curve was
presented, which resulted from the $\su(2,2)$ with representation
${\bf F}'$. The asymptotics of the quasi-momenta for these are
distinct from the ones of the string sigma-model. It may be of
insterest to understand the relevance of this spin-chain, which is
obtained from the ``Beast''ly Bethe ansatz in \refs{\BSint} and thus 
seems to be most suitable for describing states with large
hyper-charge $B$. Identifying the corresponding string configurations
may elucidate the r$\hat{\hbox{o}}$le of $B$ on the string side. 

Finally, needless to mention, to complete the discussion on the $\psu(2,2|4)$ spin chain, it would be formidable to construct the
curve for the full $AdS_5\times S^5$ string, presumably relying on the
constructions in \refs{\AF}. 
Very recently an extremely interesting proposal has appeared
  \refs{\SS}, emphasizing the S-matrix aspect of integrability. It may
  well be interesting to study the properties of the full
  $\psu(2,2|4)$ curve in this light.


\vskip 1cm
\centerline{{\bf Acknowledgements}}

\noindent
I thank Gleb Arutjunov, Andrei
Onishchenko and Henning Samtleben, and in particular Niklas Beisert for useful discussions and comments. This work is partially supported by the
Deutsche Forschungsgesellschaft, the DAAD and the
European RTN Program MRTN-CT-2004-503369.


\appendix{A}{Transfer-matrices for $\gl (m|n)$}

The properties of the transfer matrices
for Lie (super-)algebras have been discussed in \refs{\KuResh, \Kulish}, in
particular the eigenvalues of the transfer matrix for Lie
super-algebras based on classical Lie algebras are determined
therein. 
Consider a $\gl (m|n)$ spin chain with representation of highest
weight $(\mu_1, \cdots, \mu_m| \nu_1,\cdots, \nu_n)$. Further define
\eqn\afus{
a(u)= {u+1\over u}\,,\qquad 
S_{l}(u)= {u+ l/2 \over u-l /2} \,.
}
Then the eigenvalue
$T(u)$ of the super-trace $t(u)$ in the {\it fundamental
representation}, \ie, the auxiliary space is chosen to be $\V_{aux}= {\bf m}|{\bf n}$, of the transfer matrix on a state
specified by the Bethe roots $v^{(k)}_i$, $k=1\cdots m+n-1$, and is
(for the distinguished Dynkin diagram)
\eqn\transfermat{\eqalign{
T(u) &= (u+ \mu_1)^L \prod_{i=1}^{J_1} a(v_i^{(1)}-u) \cr
&+ \sum_{k=2}^m
\left((u+\mu_k)^L \prod_{i=1}^{J_{k-1}} a(u-v_i^{(k-1)})
\prod_{i=1}^{J_k} a(v_i^{(k)}-u) \right) \cr
&- (u-\nu_1)^L \prod_{i=1}^{J_m} a(v_i^{(m)}-u) \prod_{i=1}^{J_{m+1}}
a(u-v_i^{(m+1)}) \cr
&-  \sum_{k=2}^{n-1}
\left((u-\nu_k)^L \prod_{i=1}^{J_{m+k-1}} a(v_i^{(m+k-1)}-u)
\prod_{i=1}^{J_{m+k}} a(u-v_i^{(m+k)}) \right) \cr
&- (u-\nu_n)^L \prod_{i=1}^{J_{m+n-1}} a(v_i^{(m+n-1)}-u) \,.
}}
The Bethe equations for this algebra read
\eqn\bethegl{\eqalign{
(S_{1}(v_{j}^{(1)}))^L &= 
\prod_{i=1, i\not= j}^{J_1} S_2 (v^{(1)}_j - v_i^{(1)}) 
\prod_{i=1}^{J_2} S_{-1} (v_j^{(1)}- v_i^{(2)}) \cr
1&= 
\prod_{i=1}^{J_1} S_{-1}(v^{(2)}_j- v_i^{(1)}) 
\prod_{i=1, i\not= j}^{N_2} S_2 (v^{(2)}_j - v_i^{(2)}) 
\prod_{i=1}^{J_3} S_{-1} (v_j^{(2)}- v_i^{(3)}) \cr
& \quad \cdots \cr
1&= 
\prod_{i=1}^{J_{n-1}} S_{-1}(v^{(n-1)}_j- v_i^{(n-1)})
\prod_{i=1}^{J_{n+1}} S_{1}(v^{(n)}_j- v_i^{(n+1)}) \cr
1&= 
\prod_{i=1}^{J_{n+1}} S_{-1}(v^{(n+1)}_j- v_i^{(n)}) 
\prod_{i=1, i\not= j}^{N_{n+1}} S_2 (v^{(n+1)}_j - v_i^{(n+1)}) 
\prod_{i=1}^{J_{n+2}} S_{-1} (v_j^{(n+1)}- v_i^{(n+2)})\cr
& \quad \cdots \cr
1&= 
\prod_{i=1}^{J_{n+m-2}} S_{-1}(v^{(n+m-1)}_j- v_i^{(n+m-2)}) 
\prod_{i=1, i\not= j}^{J_{n+m-1}} S_2 (v^{(n+m-1)}_j - v_i^{(n+m-1)}) \,.
}}

The conventions are mapped to the ones used standard-wise in the
applications to gauge theory by replacing
\eqn\conventions{
u \rightarrow -i u\,,\qquad v^{(1)}_i \rightarrow u_i= -i (v^{(1)}_i + \half)\,.
}
{\it E.g.} then the eigenvalue of the transfer matrix \transfermat\
reduces for $\su(2)$ to the standard form
\eqn\sutwoT{
T(u) /(u+1)^L \rightarrow \prod_{k}^{J_1} {u-u_k -{i\over 2}\over u-u_k
+{i\over 2}} + \left({u\over u+i}\right)^L  
\prod_{k}^{J_1} {u-u_k +{3i\over 2}\over u-u_k +{i\over 2}} \,.
}
Similarly, for $\su(4)$ it yields the expressions 
\eqn\sufourT{\eqalign{
T(u)/ (u+1)^L   \rightarrow 
& \prod^{J_1} {u-u^{(1)}_i - {3 i\over 2} \over u-u^{(1)}_i - {i\over 2}}\cr
+ \left({u\over u+i}\right)^L &\prod^{J_1} {u-u^{(1)}_i + { i\over 2} \over u-u^{(1)}_i - {i\over 2}} 
\prod^{J_2} {u-u^{(2)}_i - i \over u-u^{(2)}_i }  \cr
+\left({u\over u+i}\right)^L
&\prod^{J_2} {u-u^{(2)}_i + i \over u-u^{(2)}_i } 
\prod^{J_3} {u-u^{(3)}_i -{ i\over 2} \over u-u^{(3)}_i + {i\over
2}} \cr
+\left({u\over u+i}\right)^L 
&\prod^{J_3} {u-u^{(3)}_i +{ 3i\over 2} \over u-u^{(3)}_i + {i\over 2}} \,.
}}
for the maps $u \rightarrow -i u$ as well as
\eqn\conventions{
v^{(1)}_i  \rightarrow u^{(1)}_i=-i (v^{(1)}_i - \half)
\,,\quad 
v^{(2)}_i \rightarrow  u^{(2)}_i= -i (v^{(2)}_i + 0)
\,,\quad 
v^{(3)}_i \rightarrow  u^{(3)}_i=-i (v^{(3)}_i + \half)
\,.
}

Finally, for $\ufrak (2,2|4)=\gl(4|4)$ the transfer matrix eigenvalue
in \transfermat\ result for the slightly different choice of Cartan
matrix (corresponding to the distinguished `Beastly' Dynkin diagram) for the $\slfrak(4|4)$ sub-super-algebra 
\eqn\CMalt{
M'=
\pmatrix{
2& -1 &    & & & &  \cr
-1 & 2 & -1 & & & &  \cr
  & -1 & 2 &-1 & & &\cr
  &  & -1 & 0& 1& & \cr
  &  & &1 &-2 &1 & \cr
  &  & & &1 & -2 &1  \cr
  &  & & & & 1 & -2  
}}
as
\eqn\glTtildemat{\eqalign{
T'(u) 
&= (u+\mu_1)^L {R_1(u-i)\over R_1(u)} \cr  
& +(u+\mu_2)^L {R_1(u+i)\over R_1(u)} {R_2(u-i)\over R_2(u)}   
  +(u+\mu_3)^L {R_2(u+i)\over R_2(u)} {R_3(u-i)\over R_3(u)}   \cr 
& +(u+\mu_4)^L {R_3(u+i)\over R_3(u)} {R_4(u-i)\over R_4(u)}   
  -(u-\mu_5)^L {R_4(u-i)\over R_4(u)} {R_5(u+i)\over R_5(u)}   \cr
& -(u-\mu_6)^L {R_5(u-i)\over R_5(u)} {R_6(u+i)\over R_6(u)}   
  -(u-\mu_7)^L {R_6(u-i)\over R_6(u)} {R_7(u+i)\over R_7(u)}   \cr
& -(u-\mu_8)^L {R_7(u-i)\over R_7(u)} \,.
}}
Finally, the quasi-momenta for this case are computed. In the thermodynamic limit
$L\rightarrow\infty$ the transfer matrix takes the from $T= \sum_k
\epsilon_k e^{i\p_k}$, with signs $\epsilon_k$ depending on the chosen
Dynkin diagram. From \Rlimit\ the quasi-momenta can be read off as
\eqn\tildep{\eqalign{
p'_1 &= +\Gt'_1(u)\cr
p'_2 &= +\Gt'_2(u) -\Gt'_1(u) \,,\quad
p'_3  = +\Gt'_3(u) -\Gt'_2(u) \,, \quad
p'_4  = +\Gt'_4(u) -\Gt'_3(u) \cr
p'_5 &= -\Gt'_5(u) +\Gt'_4(u) \,, \quad
p'_6  = -\Gt'_6(u) +\Gt'_5(u) \,, \quad
p'_7  = -\Gt'_7(u) +\Gt'_6(u) \cr
p'_8 &= -\Gt'_7(u)  \,.
}}
The distinguished Dynkin diagram is \eg, suitable for the discussion
of the $\su(2,2)$ subsector, as done in section 6, where also the singular
resolvents $\Gt'_k$ for this sector are detailed.


\appendix{B}{Algebraic Curve for the QCD Spin Chain}

In the main body of the paper we discussed the $\su(2,2)$ spin chain
and its associated algebraic curve, when the 
representation at each spin chain site has highest weight $V_{\bf
Z}=(0,-1,0)$. An alternative choice, would be to truncate the ${\bf F}$
representation of $\psu(2,2|4)$ to $\su(2,2)$. 
From the SYM point of view, this sector is built on the ground state corresponding
to $\Tr {\cal F}^L$, which 
has vacuum energy $3L$ and is distinct from the one in
section 6 (which on the contrary is half-BPS and has vanishing vacuum
energy). In fact it being a highly excited state in the SYM it
does not make sense to consider the thermodynamic limit around
it. However the discussion here may be of interest for the QCD spin
chain, where precisely this representation for $\su(2,2)$ is chosen
\refs{\BFHZ}, \ie, the vaccuum is the anti-ferromagnetic ground
state. So this appendix should be understood as constructing the algebraic
curve for the QCD spin chain (rather than that for a
subsector of ${\cal N}=4$ SYM). 

The reduction to this Lie sub-algebra is most transparent with the 
distinguished (``Beast'') choice for the Dynkin diagram of
$\psu(2,2|4)$ \refs{\BSint}. 
The highest weight of ${\bf F}$ in this case is $V_{\bf F}=
(0,-3,2;0;0,0,0)$. This appendix contains a succinct discussion of the
algebraic curve when the representation at each site is the sub-module
of ${\bf F}$, obtained by acting with $\su(2,2)$ on $V_{\bf F}$.

The relation between the excitation numbers and weights for the
distinguished Dynkin diagram were obtained in \refs{\BSint} and the
relevant parts for the present $\su(2,2)$ chain are
\eqn\suttrelF{
J'_k = \pmatrix{
& -1 + &\half (\Delta_0 -s_2) \cr
& -2 + &\Delta_0 \cr
& \ \ 0  + &\half (\Delta_0 -s_1) 
}\,,
}
where this is for states with $B=L=1$, \ie, $B$ is chosen large
before taking the thermodynamic limit. 
 
The singular resolvents and quasi-momenta are extracted from the
transfer matrix in appendix A, where all the $G'_k=0$ for $k=4\cdots
7$
\eqn\GprimeF{\eqalign{
\Gt'_1 (u) &= G'_1(u) - {1\over u} \cr
\Gt'_2 (u) & = G'_2 (u) - {2\over u} \cr
\Gt'_3(u) & = G'_3(u) \,,
}}
and
\eqn\pprimeF{\eqalign{
p'_1(u) &= \Gt'_1(u) \cr
p'_2(u) &= \Gt'_2(u) - \Gt'_1(u) \cr
p'_3(u) &= \Gt'_3(u) - \Gt'_2(u) \cr
p'_4(u) & = -\Gt'_3(u) \,.
}}
The Bethe equations in the thermodynamic limit take the form
as expected from the general considerations in \betheresolved\ with
the weight vector $V_{\bf F}'=(0,-3,2)$
\eqn\betheprimeF{\eqalign{
2 \Gslash'_1 (u) - G'_2(u) 
&=  \pslash'_1 (u) - \pslash'_2 (u) - {V'_{j_1} \over u}
=2\pi n_j^{(1)} - {V'_{j_1} \over
u}\,, \qquad  u\in \C_j^{(1)} \cr
- G'_1(u) + 2 \Gslash'_2 (u) - G'_3(u) 
&=  \pslash'_2 (u) - \pslash'_3 (u) - {V'_{j_2} \over u}
=2\pi n_j^{(2)} - {V'_{j_2} \over
u}\,, \qquad  u\in \C_j^{(2)} \cr
2 \Gslash'_3 (u) - G'_2(u) 
&=  \pslash'_3 (u) - \pslash'_4 (u) - {V'_{j_3} \over u}
=2\pi n_j^{(1)} - {V'_{j_3} \over
u}\,, \qquad  u\in \C_j^{(3)}  \,,
}}
Invoking \Gasymp\ and \suttrelF\ the asymptotics at $u=\infty$ are
found to be
\eqn\pprimeasymp{\eqalign{
p'_1(u) &=  {1\over u} \left(+ \half \Delta_0 -\half s_2  \right) +
O\left({1\over u^2}\right) \cr 
p'_2(u) &=  {1\over u} \left(+ \half\Delta_0+ \half s_2 \right) +
O\left({1\over u^2}\right)\cr 
p'_3(u) &=  {1\over u} \left(- \half \Delta_0 -\half s_1  \right) +
O\left({1\over u^2}\right) \cr 
p'_4(u) &=  {1\over u} \left(- \half\Delta_0+ \half s_1  \right) +
O\left({1\over u^2}\right) \,.
}}
The asymptotics at $u=0$ are obtained from \Tasymp. At finite $L$ the
expansion is
\eqn\Tprimeasymp{
T_{\bf F} (u+i) = 
{R_2(u+ {3\over 2}i )\over R_2(u- {3\over 2 }i)}  
{R_3(u-i)\over R_3(u+i)} + O(u^L)\,,
}
which for $L\rightarrow \infty$ limits to 
\eqn\Tprimelim{
T_{\bf F} (u+i) \rightarrow \exp \left(i (-3 \Gt'_2(u) + 2
\Gt'_3(u))\right) \,.
}
Thus at $u=0$ the combination of resolvents 
\eqn\localcharges{
-3\Gt'_2(u) + 2 \Gt'_3(u) =
-p_1'(u) - p_2'(u) + 2 p_3'(u) =
 {6\over u} +
\sum_{r=0}^\infty u^r Q_r \,. } 
gives rise to the local
charges $Q_r$. The asymptotics of the quasi-momenta at zero are
\eqn\pprimezeroF{
p'_1(u) = -{1\over u}+ O(1)\,,\qquad 
p'_2(u) = -{1\over u}+ O(1)\,,\qquad
p'_3(u) = +{2\over u}+ O(1)\,,\qquad
p'_4(u) = 0 + O(1)\,.
}
Note that the asymptotics at $u=\infty$ agree with the ones for the
representation ${\bf Z}$ (up to an overall sign),
however the ones at $u=0$ \pprimezero\ and
\pprimezeroF\ are distinct. 

The curve for this subsector is written in terms of $y_k = u^2 {d^2
p_k\over du^2}$ as
\eqn\curveprime{
\sum_{i=0, i\not= 3}^4 P_i(u) y^i = P_4 (u) \prod_{k=1}^4 (y- y_k(u))
=0 \,.
}
The moduli-count is identical to the one for the curve in section
6. However, due to the different asymptotics of the quasi-momenta, the
curves are distinct. It may be interesting to determine the dual
string configuration to the spin-chain discussed in this appendix. 


\listrefs
\bye